\documentstyle[epsfig,times]{mn}
\def\spose#1{\hbox to 0pt{#1\hss}}

\def\lta{\mathrel{\spose{\lower 3pt\hbox{$\mathchar"218$}}\raise 2.0pt\hbox{$\mathchar"13C$}}}
\def\gta{\mathrel{\spose{\lower 3pt\hbox{$\mathchar"218$}}\raise 2.0pt\hbox{$\mathchar"13E$}}}
\def\arcsec{$^{\prime\prime}$}

\title[HST and UKIRT imaging observations - II.]
{HST and UKIRT imaging observations of $\mathbf{z \sim 1}$ 6C radio
galaxies - II. Galaxy morphologies and the alignment effect}

\author[K.\,J.\, Inskip {\it et al.}]
{K.\,J.\, Inskip$^{1,2}$\footnotemark, P.\,N.\,Best$^3$,
  M.\,S.\,Longair$^2$ and H.\,J.\,A. R\"{o}ttgering$^4$\\ 
$^1$ Department of Physics \& Astronomy, University of Sheffield,
Sheffield S3 7RH\\ $^2$
Cavendish Laboratory, Madingley Road, Cambridge, CB3 0HE,\\ $^3$ 
Institute for Astronomy, Royal Observatory Edinburgh, Blackford Hill,
Edinburgh, EH9 3HJ\\ $^4$ Sterrewacht Leiden, Postbus
9513, 2300 RA Leiden, the Netherlands}

\date{}

\pagerange{\pageref{firstpage}--\pageref{lastpage}}

\pubyear{2005}

\begin{document}
\label{firstpage}
\maketitle

\begin{abstract}Powerful radio galaxies often display enhanced optical/UV 
emission regions, elongated and aligned with the radio jet
axis.  The aim of this series of papers is to separately investigate the effects
of radio power and redshift on the alignment effect, together with other radio
galaxy properties.  In this second paper, we present a deeper
analysis of the morphological properties of these systems, including
both the host galaxies and their surrounding aligned emission.

The host galaxies of our 6C subsample are well described as de
Vaucouleurs ellipticals, with typical scale sizes of $\sim 10$kpc. This
is comparable to the host galaxies of low$-z$ radio sources of similar
powers, and also the more powerful 3CR sources at the same
redshift.  The contribution of nuclear point source emission is also
comparable, regardless of radio power.

The 6C alignment effect is remarkably similar
to that seen around more powerful 3CR sources at the same redshift in terms of extent and
degree of alignment with the radio source axis, although it is
generally less luminous.    The bright, knotty features observed in
the case of the $z \sim 1$ 3CR sources are far less frequent in our 6C
subsample; neither do we observe such strong evidence for evolution in
the strength of the alignment effect with radio source size/age.
However, we do find a very strong link between the most extreme
alignment effects and emission line region properties indicative of
shocks, regardless of source size/age or power.
In general, the 6C alignment effect is still
considerably stronger than that seen around lower redshift galaxies of
similar radio powers.  Cosmic epoch is clearly just as important a
factor as radio power: although aligned emission is observed on
smaller scales at lower redshifts, the processes which produce the
most extreme features simply no longer occur, suggesting considerable
evolution in the properties of the extended haloes
surrounding the radio source.
\end{abstract}


\section{Introduction}
\footnotetext{E-mail: k.inskip@shef.ac.uk}

Powerful radio galaxies at $z > 0.3$ are often observed to be
surrounded by extensive regions of UV/optical continuum and line
emission. The excess continuum emission is generally well aligned with the radio
source axis at higher redshifts ($z \gta 0.6$), and is known as the
{\it alignment effect} (e.g. Chambers, Miley \& van Breugel 1987;
McCarthy et al 1987).  

The extended emission regions responsible for the alignment effect
have been well studied in the case of powerful radio galaxies selected
from the 3CR sample 
(e.g. McCarthy, Spinrad \& van Breugel 1995; Best, Longair \&
R\"{o}ttgering 1997). Whilst a wide range of different features are
observed for the extended emission surrounding this sample of radio
galaxies, several clear trends are generally observed. At higher
redshifts ($z \sim 1$), the optical/UV emission is typically brighter
than at lower redshifts, and can be more luminous than the host
galaxy in some cases. This emission is also typically more extensive,
with projected linear sizes of up to a few hundred kpc, and is far
more closely aligned with the radio source axis than is the case for
lower redshift systems. The emission surrounding the less powerful 3CR
radio sources at lower redshifts is rarely as extensive, and with the
exception of ionization cones produced by the powerful obscured AGN,
does not usually display any tendency for alignment with the radio
axis (Allen {\it et al} 2002).  For the high redshift 3CR sources, the
alignment effect also 
displays a strong trend with radio source size; smaller sources
typically display the most extensive, luminous aligned emission
regions (Best, Longair \& R\"{o}ttgering 1996).

Multiwavelength observations have shown that a number of different
emission mechanisms are responsible for this extended excess
UV/optical emission. These include extended line emission and nebular
continuum radiation (Dickson et al 1995), scattering of the UV
continuum from the AGN (e.g. Tadhunter et al 1992; Cimatti et al 1993)
and young stars produced in a radio jet induced starburst (McCarthy et
al 1987). The relative contributions of each of these processes vary
from source to source, and generally the excess emission cannot be
accounted for by any single mechanism. It is clear from the
observed trends that not only do different processes vary in
importance at different stages of the life of a given radio source,
but that the mechanisms which produce the excess emission also vary
with redshift. However, it is not obvious whether this evolution in
the properties of the emission regions surrounding the 3CR sources
reflects a genuine evolutionary trend with redshift, or if it is
instead due to the decreasing radio power of the lower redshift
systems. With such strong apparent links between the radio source and
the aligned/excess emission, both in terms of the morphology and
emission properties of the alignment effect, the necessity for
breaking this degeneracy is clear.

The 6C sample provides an ideal population of radio galaxies which, in
conjunction with the well-studied 3CR sample, can be used to break
the degeneracy between redshift and radio source power present in
studies of a single flux limited radio galaxy sample.  At $z \sim
1$ the 6C radio sources are typically $\sim 6$ times less powerful
than the 3CR sources at the same redshift, and matched in radio power
to 3CR sources at lower redshifts of $z \sim 0.2-0.5$.
Deep spectroscopic observations of $z \sim 1$ 6C sources have enabled
the importance of cosmic epoch and the radio source parameters on the
properties of the extended emission line structures to be determined
independently  (Inskip {\it et al}
2002b; Inskip {\it et al} 2002c).  These studies were able to
demonstrate that both radio power and redshift were independently
important for the kinematic properties of the gas, but that ionization
state only depended on the radio source parameters and not on redshift.
We now intend to use the same samples to address the question of
the evolution of the morphological properties of these systems, as
many of the mechanisms by which aligned continuum 
emission can be produced are also highly 
dependent on the properties of the radio source and AGN. 
 
In order to correctly interpret the properties of the aligned emission
regions, it is also necessary to have a thorough understanding of the host galaxy
morphologies. It is therefore important to determine the structural
parameters of the host galaxies
precisely, particularly with regard to the presence of an unresolved
point source contribution from an incompletely obscured AGN.  For the
later consideration of galaxy colours, this AGN contribution must be
included in the excess UV emission, which does not necessarily
originate wholly from any external alignment effect.

At redshifts of up to at least $z \sim 1$, the host galaxies of the
powerful 3CR radio sources are usually found to be giant ellipticals
with characteristic sizes of $\sim$ 10-12kpc (BLR98; McLure \& Dunlop
2000),  and are amongst the most massive galaxies known at high 
redshifts.
The stellar populations of these galaxies are also
amongst the oldest found; the $K$ band magnitude-redshift relation for
the most powerful 3CR radio 
sources is consistent with passively evolving galaxies with old
stellar populations formed at $z \sim 5$ or earlier.  More
varied morphologies are observed at higher redshifts where the host
galaxies may still be in the process of formation (e.g. Pentericci et
al 1999).  
The 6C radio galaxy sample, selected at a lower limiting radio flux
density than the 3CR sample, displays a similar $K-z$ relation at
low$-z$.  However, beyond $z \sim 0.8$, the $K$
magnitudes of these sources are typically $\sim 0.6$ magnitudes
fainter (Eales \& Rawlings 1996), a variation between
the samples which can be interpreted in several ways (Inskip et al
2002a).  The more luminous 
galaxies could have a greater AGN contamination in their $K$
band magnitudes from direct nuclear light, emission lines or other AGN
induced emission; the stellar populations of the different samples
could differ in age or metallicity; or more simply, the less powerful high$-z$
radio sources could be hosted by less massive galaxies.  
A mass difference can be easily understood:
if the radio sources are being fuelled at the Eddington limit
then the radio power depends upon black hole mass, and it is now well
established that black hole mass and galaxy mass are correlated
(Kormendy \& Richstone 1995).  
An early study of the host galaxies of $z \sim 1$ 6C sources
(Roche, Eales \& Rawlings 1998, hereafter RER98) suggested that are
typically hosted by elliptical galaxies with a far smaller average physical
size than 3CR sources at the same redshift.  However, as this result was based upon measurements
of galaxy sizes $<$ 1\arcsec with 0.5\arcsec pixels in 1.5\arcsec
seeing conditions, confirmation of this result using improved
instrumentation was certainly desirable.
 
In the first paper in this series (Inskip et al 2003, {\it hereafter Paper I}), we presented
the results of HST and UKIRT imaging observations of a complete sample
of $z \sim 1$ 6C radio galaxies. Full details of the sample selection and
observations are provided within that paper.   In this paper, we study
the morphologies of the host galaxies and the aligned emission regions in greater depth.  
The results of 2-d modelling of the
$K-$band morphologies of the host galaxies are presented in section 2,
through which we have deduced the structural parameters for the host
galaxies, including any contribution from an incompletely obscured
nuclear point source component. The resulting radial
profiles, characteristic radii, ellipticity and  percentage
contribution of unresolved nuclear emission are contrasted with the
results of other radio galaxy samples. In section 3, we quantify various
properties of the aligned emission on the basis of the elongation and
orientation of the rest-frame UV aligned structures visible in our HST
observations.  The properties
of the 6C subsample are compared with the 
data for the matched 3CR sample at the same redshift. The behaviour
of the aligned emission is studied as a function of both radio source
size and redshift, as well as over the factor of $\sim$six difference in
radio power between the two samples.  In section 4, we discuss the
range of different morphologies observed in the two samples, and the
implications of these combined results.   
In the third paper in this series, we will present a detailed study of
the galaxy colours for both $z \sim 1$ subsamples.  Cosmological parameters of $\Omega_0=0.3$,
$\Omega_\Lambda=0.7$ and $H_{0}=65\,\rm{km\,s^{-1}\,Mpc^{-1}}$ are
assumed throughout this paper.

\section{Galaxy morphologies}
\subsection{Properties of the host galaxies}
Our deep K-band observations of the 6C sources (with a limiting
1-sigma magnitude of $K \sim 23.4$ per square arcsec) are particularly well 
suited to the study of the host galaxy morphologies, through the
fitting of their radial/surface profiles.  In order to carry out this
fitting, it was first necessary to obtain accurate point source
profiles for each of the fields. The observations for each individual
source were carried out over several nights under quite variable
seeing conditions (ranging from 0.4\arcsec to 1.2\arcsec; see Paper I
for full details).    The point spread functions for each source are
therefore quite complex, and are best reproduced by extracting 2-d 
profiles for all stars in the field of each source, normalising to
unit flux and taking the average profile. As a guideline, the average
seeing for each source ranges from 0.6\arcsec to 0.9\arcsec, with an
mean value for the sample as a whole of 0.75\arcsec.
In order to accurately model the galaxies, nearby objects were masked
where these 
were well separated from the 
host galaxy; where this was not possible (e.g. 6C1100+35, 6C1129+37
and 6C1256+36) a second galaxy and/or point source is included in our 
modelling to account for the extra flux from the secondary object.
Galaxy models were then
convolved with these point spread functions and used to fit
the surface profiles of the galaxies, using available least squares
minimisation 
IDL routines. 
In particular, we have made use of the IDL (Interactive Data Language)
routine   
\textsc{mp2dfunfit.pro}, part of the \textsc{mpfit} IDL
package\footnote{a non-linear least squares curve fitting package available via
http://astrog.physics.wisc.edu/$\sim$craigm/idl/fitting.html developed
by Craig Markwardt.}.
The \textsc{mpfit} package iteratively searches for the best fit model
parameters via minimisation of the weighted squared difference between
the data and model.  This least-squares minimisation uses the
Levenberg-Marquardt technique (More, 1977), which numerically
calculates the derivatives of the assumed function/model via a finite
difference approximation (for further details, see More \& Wright
(1993) and MINPACK-1, 
available from netlib at http://www.netlib.org).
In order to weight the data at different radii uniformly, rather than
giving preference to either the higher signal to noise data at the
centre of each galaxy or the lower signal to noise data at larger radii
for which we have more pixels, we have weighted each pixel by $1/2\pi
r$, normalising the weights to a mean value of 1.0 so as not to alter
the resulting value of $\chi^2$ produced by the code. We have
determined that this weighting scheme does not introduce any strong
biases into the fitting
procedure.
Formal 1-sigma uncertainties on each parameter are computed from the
square root of the diagonal elements of the covariance
matrix for the parameters being fitted; these errors are obtained from
the code via the
inclusion of an error array, consisting of the measured RMS noise of
the data divided by the pixel weights.

\begin{table*}
\begin{center} 
\caption{Results of galaxy surface profile fitting. Source name and
  redshift are listed in columns 1 \& 2. The filter used for each
  observation is listed in column 3. The best fit
  effective radius for each source is given in columns 4 (arcsec) and
  5 (kpc), with the
  best-fit nuclear point source contribution listed in column 6.
  Column 7 gives the value of the reduced $\chi^2$ for the fit.}
\begin{tabular} {cccr@{$\pm$}lr@{$\pm$}lcr@{$\pm$}lc} 
Source  &Redshift & Filter &\multicolumn{2}{c}{ $r_{\rm eff}$ (arcsec)}&\multicolumn{2}{c}{ $r_{\rm eff}$ (kpc)} & 
\multicolumn{3}{c}{Point source fraction}& {Reduced
  $\chi^2$}\\
(1) &(2) &(3) &\multicolumn{2}{c}{(4)}&\multicolumn{2}{c}{(5)} & 
\multicolumn{3}{c}{(6)}& {(7)}\\
\hline
6C0825+34 &1.467& K  &\multicolumn{7}{c}{S/N too low for accurate fitting$^1$}&-- \\\vspace{5pt}
&& F814W &\multicolumn{2}{c}{--}&\multicolumn{2}{c}{--}&&\multicolumn{2}{c}{--}&-- \\
6C0943+39 &1.035& K  &1.90&$0.40^{\prime\prime}$&16.5 &3.5 kpc&&41.5&3.5\% & 0.98\\\vspace{5pt}
& & F702W & \multicolumn{2}{c}{{\it
    1.9\arcsec}$^2$}&\multicolumn{2}{c}{{\it 16.5 kpc}$^2$}& & 36 & $^6_4$\% &1.16 \\	
6C1011+36 &1.042& K  &1.00&$^{0.19}_{0.09}$$^{\prime\prime}$&
8.7&$^{1.7}_{0.8}$ kpc&&   28.8&1.0\%  &1.10\\\vspace{5pt}
& & F702W & 1.04&0.05\arcsec&9.0&0.4 kpc& & 25 & 2\% &1.27 \\	
6C1017+37 &1.053& K  &0.20&0$.05^{\prime\prime}$&1.7&0.4 kpc&&  4&$^{13}_{4}$\% & 1.09\\\vspace{5pt}
& & F702W & 0.28&0.1\arcsec&2.4&0.9 kpc& & 12 & 3\% &2.17 \\	
6C1019+39$^{3}$ &0.922& K & 1.18&$0.01^{\prime\prime}$&10.0&0.1 kpc&& 0.0 & 1.0\%&1.25\\	
& & F814W & 1.04&0.06\arcsec&8.8&0.5 kpc& & 0 & 1\% &1.29 \\\vspace{5pt}              	
& & F606W & 1.16&0.2\arcsec&9.8&1.7 kpc& & 0 & 2\% &1.20 \\	
6C1100+35 &1.440& K & 0.16&$0.04^{\prime\prime}$ &1.5&0.4 kpc& &15 & 14\% &1.49\\\vspace{5pt}
& & F814W & 0.2&0.1\arcsec&1.9&0.9 kpc& &41  &$^{11}_{1}$\% & 1.16\\          	
6C1129+37 (Radio Galaxy)  &1.060& K &1.8&$0.2^{\prime\prime}$&15.7&1.7
kpc&& 18.0&1.0\%& 0.97\\
 (Companion Galaxy)&& K & 1.7&$0.2^{\prime\prime}$ &14.9& 1.7 kpc &&
14.0&1.0\%  &--\\\vspace{5pt}
(Both Galaxies) && F702W &\multicolumn{7}{c}{UV morphology
too complex for galaxy fitting.}&--\\ 
6C1204+35&1.376& K  & 0.85&$0.10^{\prime\prime}$ &7.7&0.9 kpc && 13.0&1.0\% & 1.00\\\vspace{5pt}   
& & F814W & \multicolumn{7}{c}{Strong dust lane prevents accurate
  fitting of galaxy.}&--\\      	
6C1217+36&1.088& K  & 2.50& $0.15^{\prime\prime}$ &22.0&1.3 kpc&&  25.0&0.5\% & 1.32\\
& & F814W & \multicolumn{2}{c}{{\it 2.5\arcsec}$^4$}&\multicolumn{2}{c}{{\it 22.0 kpc}$^4$}& &  23&5\% &  1.07\\\vspace{5pt}              	
& & F606W &\multicolumn{2}{c}{{\it 2.5\arcsec}$^4$}
&\multicolumn{2}{c}{{\it 22.0 kpc}$^4$}& &\multicolumn{2}{c}{$>65$\%$^5$} & -- \\	
6C1256+36 (Radio Galaxy)&1.128& K  & 0.95&$0.20^{\prime\prime}$ &8.4& 1.8 kpc&& 0.0&3.0\% &1.17\\ 
(Unresolved Companion)$^6$&& K  & \multicolumn{7}{c}{Accounts for 25\%
  of aperture flux; i.e. 34\% of
radio galaxy flux}& --\\   
(Radio Galaxy)$^7$&& F702W &\multicolumn{2}{c}{{\it 0.95\arcsec}}
&\multicolumn{2}{c}{{\it 8.4 kpc}}&&\multicolumn{2}{c}{0\%}  & 1.21 \\\vspace{5pt}
(Unresolved Companion) && F702W &\multicolumn{7}{c}{Accounts for 17\%
  of aperture flux; i.e.  20\% of
radio galaxy flux.}&--\\
6C1257+36 &1.004& K & 1.39&0.07\arcsec&12.0&0.6 kpc&&  16.0&0.5\% & 1.32\normalsize{}\\ 
& & F814W & {\it 1.39}&$^{0.05}_{0.7}$\arcsec$^8$&{\it
  12.0}&$^{0.4}_{6.0}$ kpc$^8$& & 3.5 & $^{1}_{3.5}$\% &1.29 \\       	
& & F606W & {\it 1.39}&$^{0.3}_{0.7}$\arcsec$^8$&{\it 12.0}&$^{2.6}_{6.0}$
kpc$^8$& & 11 & 3\% &1.18 \\\hline	
\multicolumn{9}{l}{Notes:} \\
\multicolumn{11}{l}{[1]: Fitting 6C0825+34 with a pure de Vaucouleurs
  profile (zero point source contribution) suggests a best fit  value
  of $r_{\rm eff} = 0.2$\arcsec.  Fitting more} \\
\multicolumn{11}{l}{  complex profiles are fully
  degenerate; neither $r_{\rm eff}$ nor the point source contribution can
  be constrained.  } \\
\multicolumn{11}{l}{[2]: Due to the extensive aligned emission
  surrounding 6C0943+39, the effective radius of this galaxy could not
  be constrained for the F702W }\\
\multicolumn{11}{l}{ filter. The value obtained from our $K$  band
  observations is assumed in order to determine the point source
  contribution in this waveband.  The 1-$\sigma$}\\
\multicolumn{11}{l}{ uncertainties on the $K$
  band effective radius have been used to help constrain the errors on
the point source contribution in the F702W emission}\\
\multicolumn{11}{l}{[3]: 6C1019+39 is the only source which has a
significant ellipticity.  The best fit value in the $K$ band is $\epsilon = 0.69$; but the
effective radius of the } \\
\multicolumn{11}{l}{  model galaxy varies very little with
  $\epsilon$.  As $\epsilon \rightarrow 0$, $r_{\rm
    eff}\rm(min)\rightarrow1.0$\arcsec. The ellipticity obtained from
  the HST images is  very similar. }\\
\multicolumn{11}{l}{[4]: 6C1217+36 has a strong nuclear point source
  contribution in both HST filters. Additionally, there is excess UV
  flux present offset slightly to }\\
\multicolumn{11}{l}{the NE from the centre of the host
  galaxy, and/or a dust lane. Due to these complicating factors, the effective radius
  cannot be constrained in the }\\ 
\multicolumn{11}{l}{HST filters.}\\
\multicolumn{11}{l}{[5]: The quoted point source contribution for
  6C1217+36 in the F606W filter is the minimum value, giving $\chi^2
  \sim 6$.}\\
\multicolumn{11}{l}{[6]: Unresolved point source offset from host
galaxy, possibly a foreground star.}\\
\multicolumn{11}{l}{[7]: S/N too low for accurate fitting; assuming the
  $K$ band value of $\rm{r_{eff}}$ suggests zero point source
  contribution to the host galaxy itself, with the }\\
\multicolumn{11}{l}{unresolved
  companion object accounting for $\sim 20$\% of the total flux. }\\
\multicolumn{11}{l}{[8]: The presence of bright excess UV emission (both to the SE of
  this galaxy and also to the NW within it) prevents good fitting of
  the effective}\\
\multicolumn{11}{l}{radius of this source in the HST filters.  The
  $K$-band value is used to constrain the nuclear point source
  contribution in these filters.}
\end{tabular}
\end{center}
\end{table*}

As was found by the earlier work of RER98, de Vaucouleurs profiles
provide a far better fit to the data than either pure point source or
exponential disk models. For those galaxies which overlap with the
RER98 sample our own modelling confirms this result, and in general 
the results of our fitting for a pure de Vaucouleurs profile produce
very similar best fit effective radii to those of RER98 (with
occasional exceptions such as 6C1129+37 which was not previously know
to consist of a pair  of elliptical galaxies).   However, a more
representative model of these galaxies should also include an
unresolved nuclear point source contribution.  A ``point source
fraction'' has been included as an additional free parameter in our
modelling of these sources, alongside the effective radius, peak
intensity, galaxy centroid and ellipticity.  
One issue with this modelling is the potential degeneracy between the 
shape of the galaxy profile and the derived nuclear point source
contribution: a combination of large $r_{\rm eff}$ and high point
source contribution often produces a similar reduced $\chi^2$ value to
a smaller galaxy with a much lower point source contribution.  This
behaviour of the reduced $\chi^2$ for various values of $r_{\rm eff}$
and the point source fraction is clearly illustrated and constrained by the shape of
our $\chi^2$ contours in figures 1-10.
We have also considered
the slightly more complex S\'{e}rsic profile (i.e. $r^{1/n}$ rather
than simply $r^{1/4}$; S\'{e}rsic 1968) for the single galaxy
systems in our sample (i.e. excluding 6C1100+35, 6C1129+37 and
6C1256+36). The S\'{e}rsic profile can account for slight variations in the
slope of the galaxy profiles compared with the predictions of a de
Vaucouleurs model.  In particular, a galaxy with a high nuclear point
source contribution may in fact be better modelled by a S\'{e}rsic
profile with $n > 4$, the steeper slope of which incorporates a
higher luminosity for the central regions. We find that for the
majority of our sources a value of $n \sim 4$ is preferred (although a
few sources are better fit by S\'{e}rsic profiles with a 
higher/lower value of $n$), confirming that in general the host
galaxies are well behaved elliptical galaxies.   However, the
uncertainties on the actual slope of the profiles are not
insignificant, and the quality of our data is not sufficient for
us to accurately constrain the profile shape (i.e. $n$), size,
point source component and other parameters independently.  Therefore,
for ease of comparison with previous datasets and given that $n \sim 4$ is
the preferred value, we restrict our detailed analysis to de
Vaucouleurs elliptical profiles with $n=4$; a
brief discussion of the effects of deviations from de Vaucouleurs profiles is
included where relevant in section 2.3. 

\begin{figure*}
\vspace{6.2 in}
\begin{center}
\includegraphics{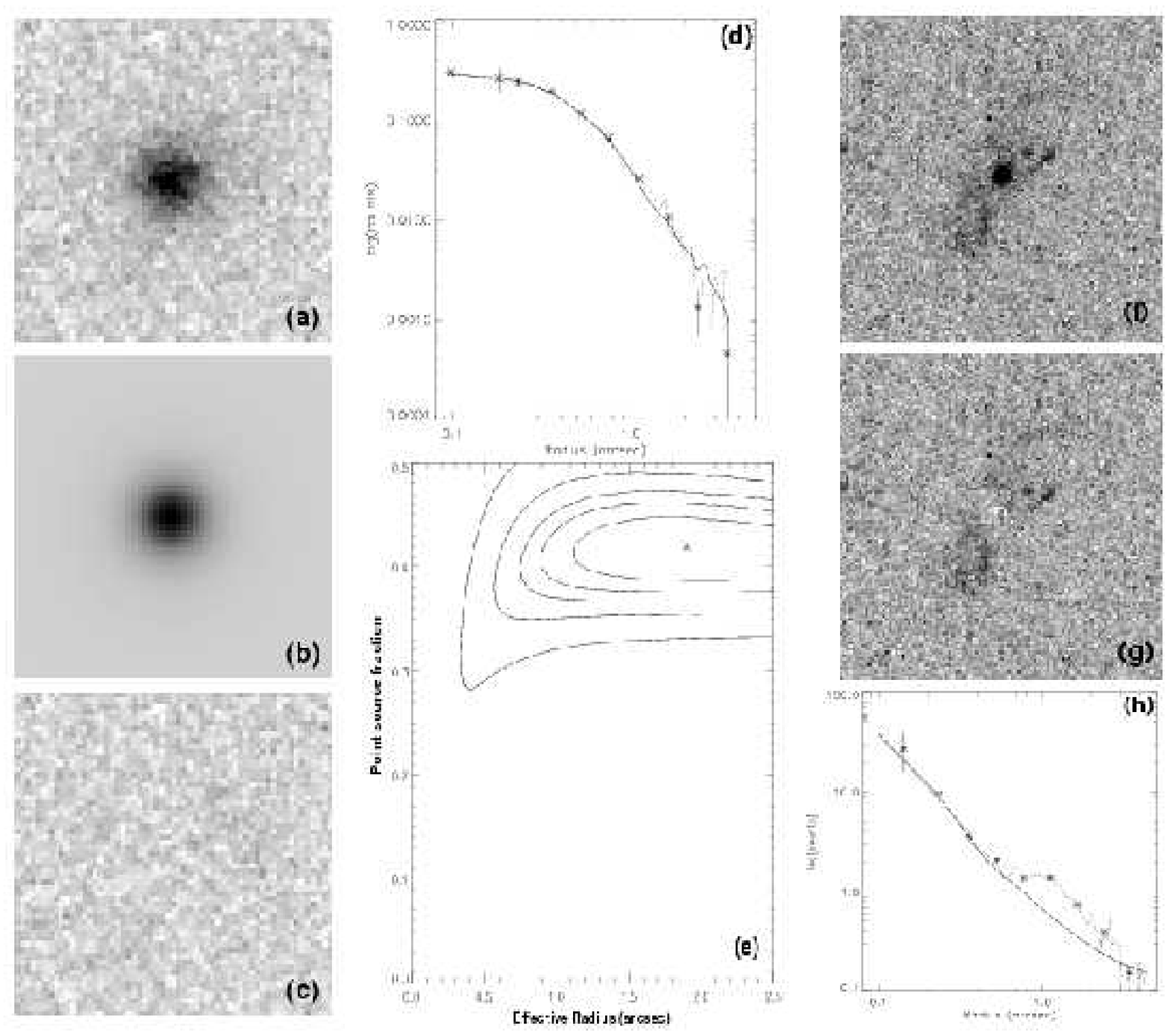}
\end{center}
\caption{Host galaxy morphology fits for the galaxy 6C0943+39. The $K-$band
  galaxy image is presented in frame (a), with the model galaxy given
  in frame (b) and the residuals after subtracting the best fitting model from
  the data displayed in frame (c).  Frame (d) illustrates a 1-d
  profile of these data (dotted line and binned points) with the best-fit model (solid line)
  overlaid. Frame (e) displays the minimum reduced $\chi^2$ (marked with a
cross), plus contours for $\chi^2_{min}
+ 1$, $\chi^2_{min} + 2$, $\chi^2_{min} + 3$, $\chi^2_{min} + 5$ \&
$\chi^2_{min} + 10$. The variation in $\chi^2$ was determined by
fitting the galaxy whilst fixing the values of $r_{\rm eff}$ and the point
source fraction at increments of 0.05 arcsec and 1\% of the total flux
respectively; other parameters (galaxy peak intensity and centroid) 
were free to vary. For models with a point source fraction of 0.0\%,
the minimum value of $\chi^2$ lies at $r_{\rm eff} \sim 0.15$\arcsec.
  Frame (f) displays the HST data for this source.  The residual flux
  after subtraction of the best fit model galaxy is displayed in frame
  (g), and frame (h) illustrates a 1-d profile of these data (dotted
  line and binned points) with the best fit model (solid line).
\label{Fig: 1}}
\end{figure*}

\begin{figure*}
\vspace{6.2 in}
\begin{center}
\includegraphics{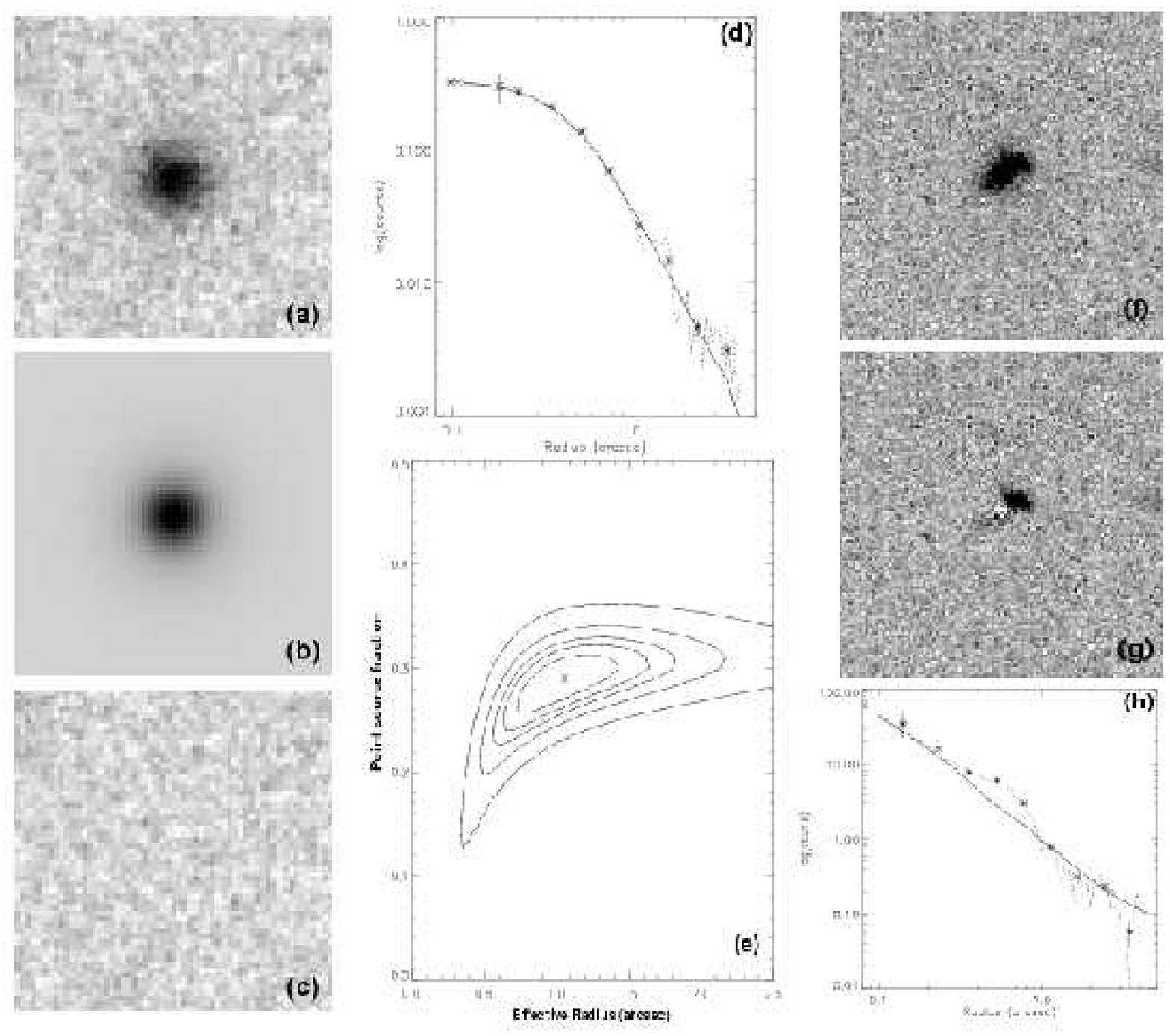}
\end{center}
\caption{Host galaxy morphology fits for the galaxy 6C1011+36. The $K-$band
  galaxy image is presented in frame (a), with the model galaxy given
  in frame (b) and the residuals after subtracting the best fitting model from
  the data displayed in frame (c).  Frame (d) illustrates a 1-d
  profile of these data (dotted line and binned points) with the best-fit model (solid line)
  overlaid. Frame (e) displays the minimum reduced $\chi^2$ (marked with a
cross), plus contours for $\chi^2_{min}
+ 1$, $\chi^2_{min} + 2$, $\chi^2_{min} + 3$, $\chi^2_{min} + 5$ \&
$\chi^2_{min} + 10$. For models with a point source fraction of 0.0\%,
the minimum value of $\chi^2$ lies at $r_{\rm eff} \sim 0.25$\arcsec.
  Frame (f) displays the HST data for this source.  The residual flux
  after subtraction of the best fit model galaxy is displayed in frame
  (g), and frame (h) illustrates a 1-d profile of these data (dotted
  line and binned points) with the best fit model (solid line).
\label{Fig: 2}}
\end{figure*}

\begin{figure*}
\vspace{6.2 in}
\begin{center}
\includegraphics{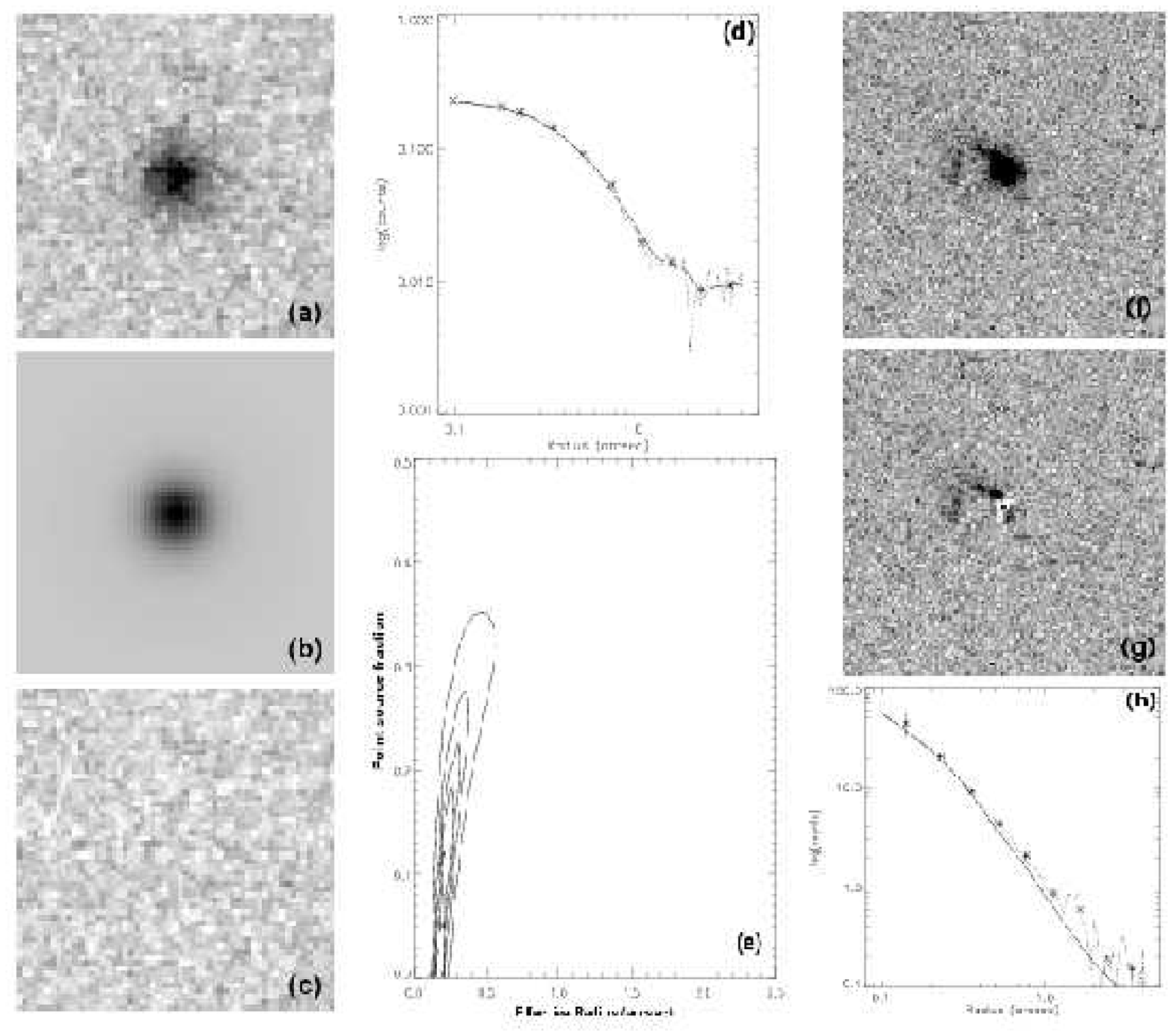}
\end{center}
\caption{Host galaxy morphology fits for the galaxy 6C1017+37. The $K-$band
  galaxy image is presented in frame (a), with the model galaxy given
  in frame (b) and the residuals after subtracting the best fitting model from
  the data displayed in frame (c).  Frame (d) illustrates a 1-d
  profile of these data (dotted line and binned points) with the best-fit model (solid line)
  overlaid. Frame (e) displays the minimum reduced $\chi^2$ (marked with a
cross), plus contours for $\chi^2_{min}
+ 1$, $\chi^2_{min} + 2$, $\chi^2_{min} + 3$, $\chi^2_{min} + 5$ \&
$\chi^2_{min} + 10$.
  Frame (f) displays the HST data for this source.  The residual flux
  after subtraction of the best fit model galaxy is displayed in frame
  (g), and frame (h) illustrates a 1-d profile of these data (dotted
  line and binned points) with the best fit model (solid line).
\label{Fig: 3}}
\end{figure*}

Tabulated results of our modelling can be found in Table 1, with errors
determined as described earlier in this section.  
Figures 1-10 display the following for each source: the K-band galaxy
images (a), best-fit model
(after convolution with the appropriate psf) (b),
the residuals obtained after subtracting this model from the original
data (c), a radially averaged 1-d profile of these data and best fit
model (d) and contours representing the reduced $\chi^2$
(ideally $\sim 1$ for an accurate fit; actual values for each source
listed in Table 1) plus 1, 2, 3, 5 and 10 (e).
Examination of the value of $\chi^2$ over the local parameter
space provides a further check on the accuracy of our fitting;
although these results do not consider the error on the other free
parameters, the $\chi^2+1$ contours cover a similar range to the
1$\sigma$ errors on each parameter deduced as part of the fitting
process, and provide a valuable visualisation of the accuracy of our
modelling.  

As a final test of the accuracy of the fitting procedures,
the parameters of
the best fitting models were used to construct artificial galaxies.
After adding noise to the data so as to reproduce the
signal--to--noise of the original observations, the same routines were
used to fit the parameters of the artificial data. In this way, we
have been able to investigate the tendency of the routine to over- or
underestimate the value of a given parameter for a particular source,
and the range of data which provide consistent results (i.e. which
produce artificial galaxies indistinguishable from the actual
object). The range of values given for errors on $r_{\rm eff}$,
$\epsilon$ and the 
point source fraction in Table 1 agree well with the range of input
data which gave 
results consistent with the observations.

\begin{figure*}
\vspace{8.2 in}
\begin{center}
\includegraphics{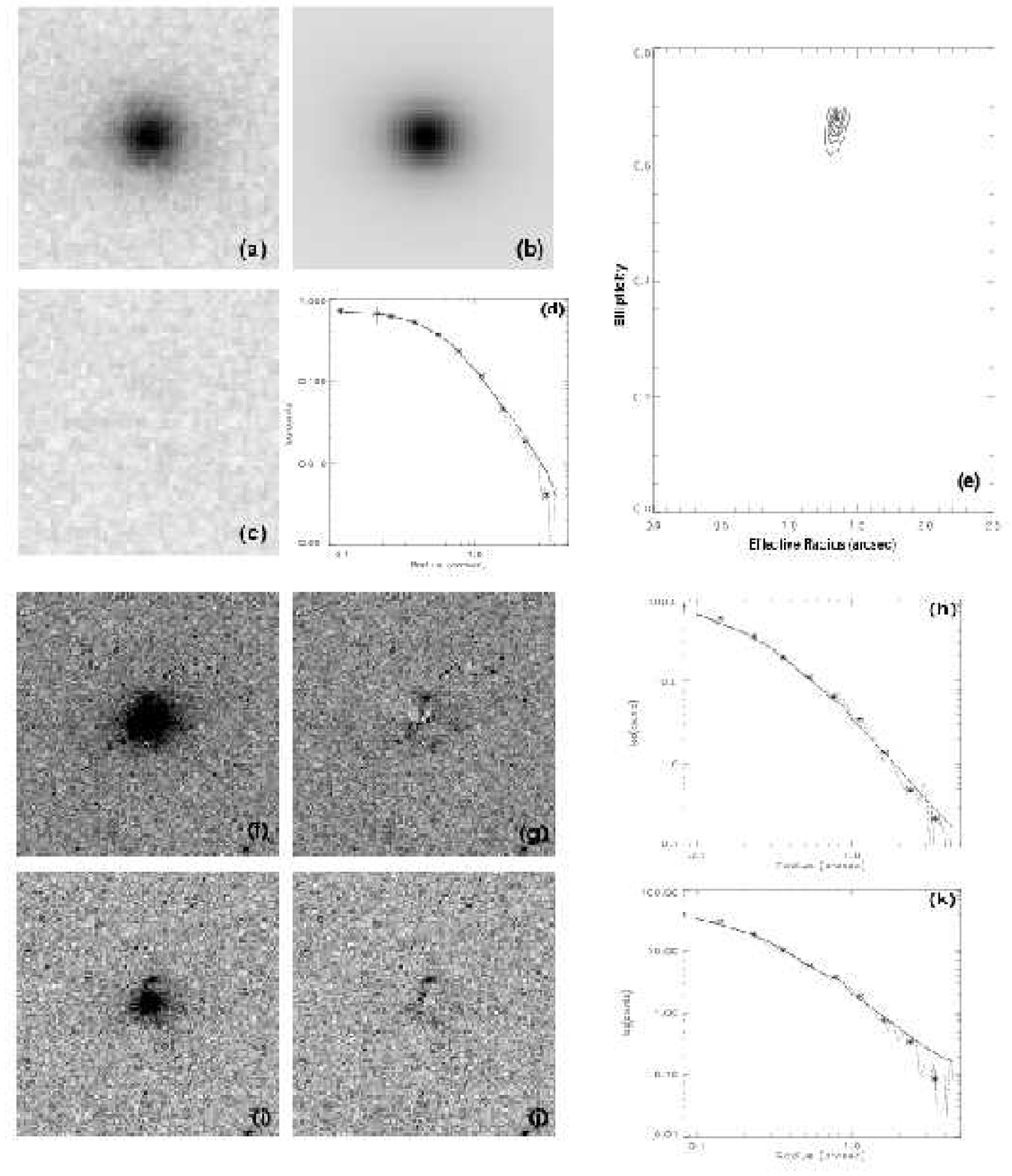}
\end{center}
\caption{Host galaxy morphology fits for the galaxy 6C1019+39. The $K-$band
  galaxy image is presented in frame (a), with the model galaxy given
  in frame (b) and the residuals after subtracting the best fitting model from
  the data displayed in frame (c).  Frame (d) illustrates a 1-d cut
  through the data (dotted line and binned points) with the best-fit model (solid line)
  overlaid. Frame (e) displays the minimum reduced $\chi^2$ (marked with a
cross), plus contours for $\chi^2_{min}
+ 1$, $\chi^2_{min} + 2$, $\chi^2_{min} + 3$, $\chi^2_{min} + 5$ \&
$\chi^2_{min} + 10$.The variation in $\chi^2$ (frame (e)) was
  determined by 
fitting the galaxy whilst fixing the values of $r_{\rm eff}$, galaxy
  ellipticity ($\epsilon$) and the point
source fraction at increments of 0.05 arcsec, 0.01 and 1\% of the total flux
respectively; other parameters (galaxy peak intensity and centroid) 
were free to vary.  As $\epsilon \rightarrow 0.0$, $r_{\rm eff} \rightarrow
  1.0$\arcsec. The point source fraction giving the minimum $\chi^2$ remains
  at $0 \pm 1$\% for all preferred values of $r_{\rm eff}$ and $\epsilon$.
  Frame (f) displays the HST data for this source in the F814W filter.  The residual flux
  after subtraction of the best fit model galaxy is displayed in frame
  (g), and frame (h) illustrates a 1-d profile of these data (dotted
  line and binned points) with the best fit model (solid line).
The same data for the F606W filter are displayed in frames (i), (j) \& (k).
\label{Fig: 4}}
\end{figure*}
\begin{figure*}
\vspace{6.2 in}
\begin{center}
\includegraphics{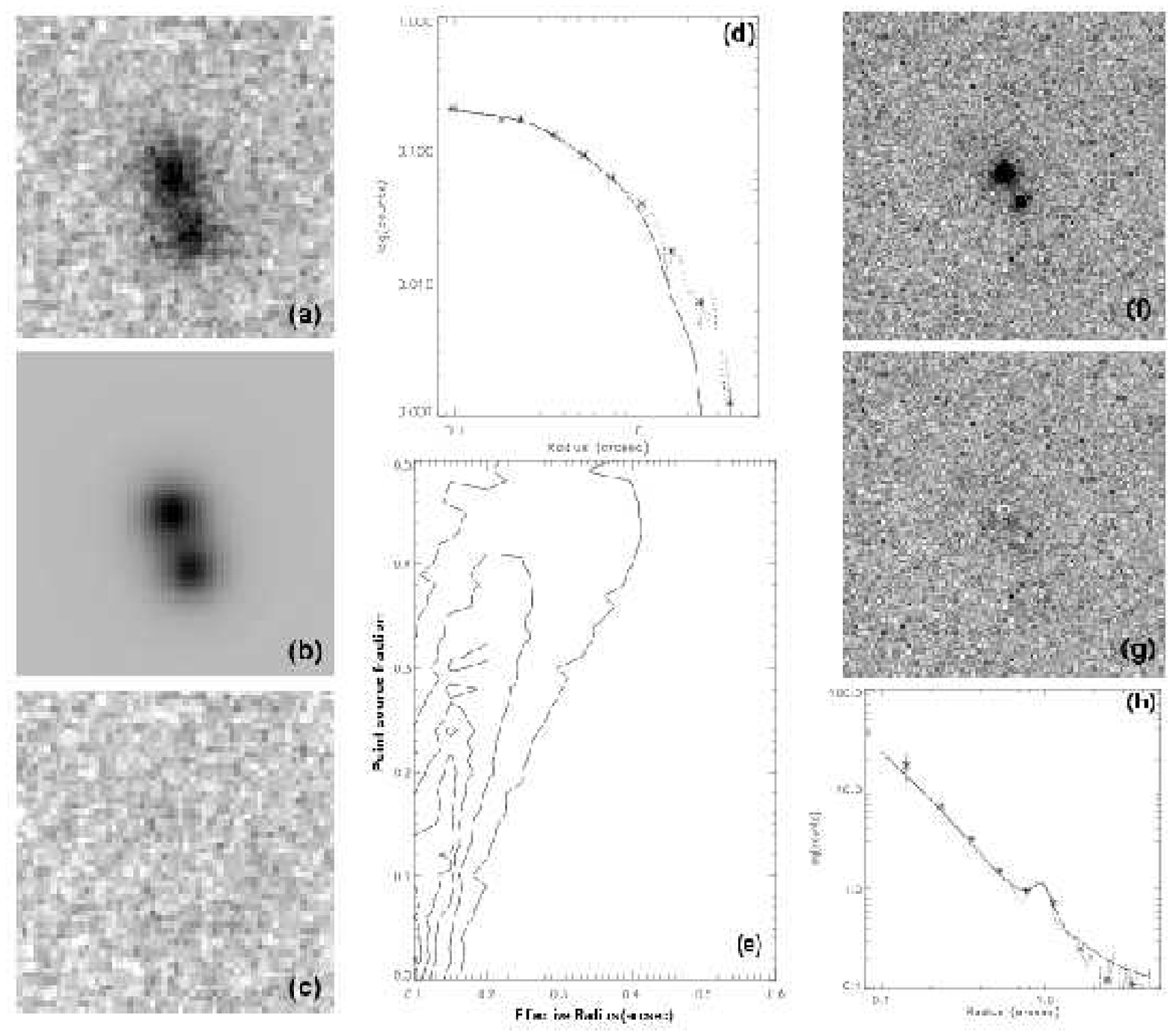}
\end{center}
\caption{Host galaxy morphology fits for the galaxy 6C1100+35. The $K-$band
  galaxy image is presented in frame (a), with the model galaxy given
  in frame (b) and the residuals after subtracting the best fitting model from
  the data displayed in frame (c).  Frame (d) illustrates a 1-d cut
  through the data (dotted line and binned points) with the best-fit model (solid line)
  overlaid. Frame (e) displays the minimum reduced $\chi^2$ (marked with a
cross), plus contours for $\chi^2_{min}
+ 1$, $\chi^2_{min} + 2$, $\chi^2_{min} + 3$, $\chi^2_{min} + 5$ \&
$\chi^2_{min} + 10$.  Due to the increases in time required
for fitting multiple sources, the variation in $\chi^2$ was initially determined at low
resolution over the same range of parameters as for other sources,
prior to being repeated over a reduced
range of values for $r_{\rm eff}$. Values of $r_{\rm eff}$ and the point
source fraction were incremented by 0.05 arcsec and 1\% of the total
  flux respectively whilst other parameters (galaxy peak intensity and
centroid, second galaxy parameters) 
were free to vary (within sensible limits in the case of the companion
object). 
  Frame (f) displays the HST data for this source.  The residual flux
  after subtraction of the best fit model galaxy is displayed in frame
  (g), and frame (h) illustrates a 1-d profile of these data (dotted
  line and binned points) with the best fit model (solid line).
\label{Fig: 5}}
\end{figure*}

\begin{figure*}
\vspace{6.2 in}
\begin{center}
\includegraphics{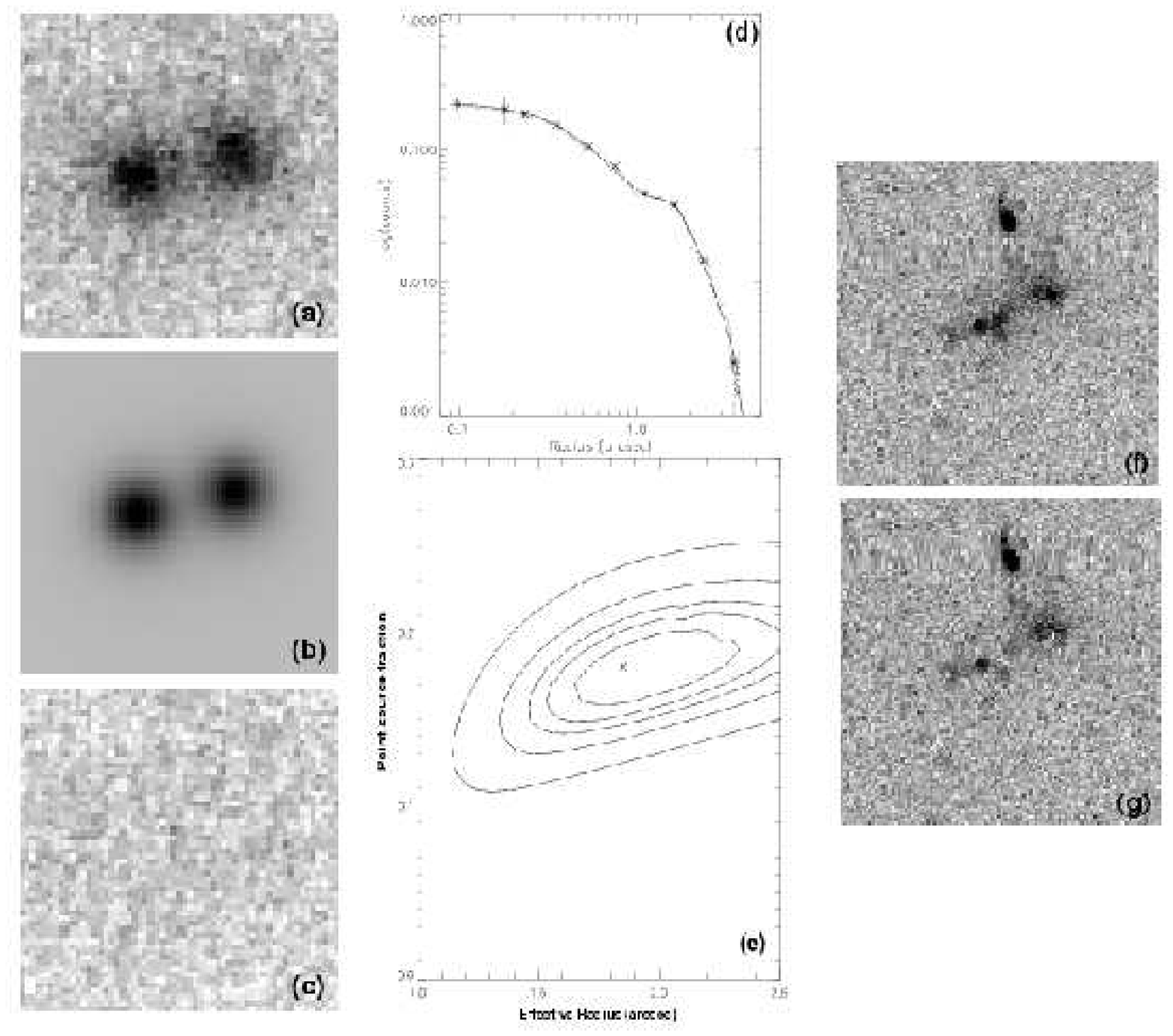}
\end{center}
\caption{Host galaxy morphology fits for the galaxy 6C1129+37. The $K-$band
  galaxy image is presented in frame (a), with the model galaxy given
  in frame (b) and the residuals after subtracting the best fitting model from
  the data displayed in frame (c).  Frame (d) illustrates a 1-d cut
  through the data (dotted line and binned points) with the best-fit model (solid line)
  overlaid. Frame (e) displays the minimum reduced $\chi^2$ (marked with a
cross), plus contours for $\chi^2_{min}
+ 1$, $\chi^2_{min} + 2$, $\chi^2_{min} + 3$, $\chi^2_{min} + 5$ \&
$\chi^2_{min} + 10$.  Due to the increases in time required
for fitting multiple sources, the variation in $\chi^2$ was initially determined at low
resolution over the same range of parameters as for other sources,
prior to being repeated over a reduced
range of values for $r_{\rm eff}$. Values of $r_{\rm eff}$ and the point
source fraction were incremented by 0.05 arcsec and 1\% of the total
  flux respectively whilst other parameters (galaxy peak intensity and
centroid, second galaxy parameters) 
were free to vary (within sensible limits in the case of the companion
object). 
  Frame (f) displays the HST data for this source.  The residual flux
  after subtraction of the best fit model galaxy is displayed in frame
  (g).
\label{Fig: 6}}
\end{figure*}

\begin{figure*}
\vspace{6.2 in}
\begin{center}
\includegraphics{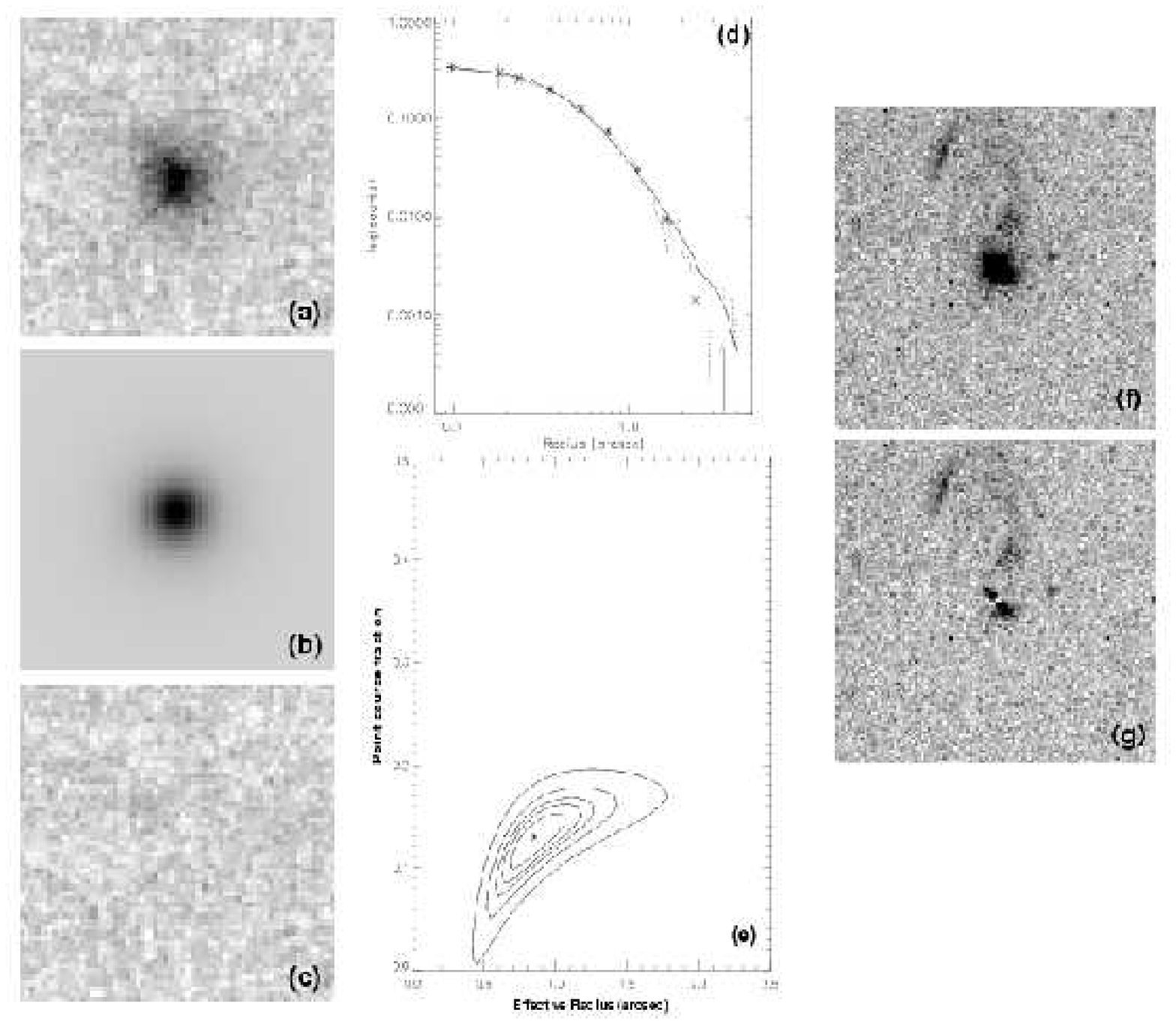}
\end{center}
\caption{Host galaxy morphology fits for the galaxy 6C1204+35. The $K-$band
  galaxy image is presented in frame (a), with the model galaxy given
  in frame (b) and the residuals after subtracting the best fitting model from
  the data displayed in frame (c).  Frame (d) illustrates a 1-d cut
  through the data (dotted line and binned points) with the best-fit model (solid line)
  overlaid. Frame (e) displays the minimum reduced $\chi^2$ (marked with a
cross), plus contours for $\chi^2_{min}
+ 1$, $\chi^2_{min} + 2$, $\chi^2_{min} + 3$, $\chi^2_{min} + 5$ \&
$\chi^2_{min} + 10$. For models with a point source fraction of 0.0\%,
  the minimum value of $\chi^2$ lies at $r_{\rm eff} \sim
  0.45$\arcsec. 
  Frame (f) displays the HST data for this source.  The residual flux
  after subtraction of the best fit model galaxy is displayed in frame
  (g). 
\label{Fig: 7}}
\end{figure*}

\subsection{Rest Frame UV morphologies}

\begin{figure*}
\vspace{7.85 in}
\begin{center}
\includegraphics{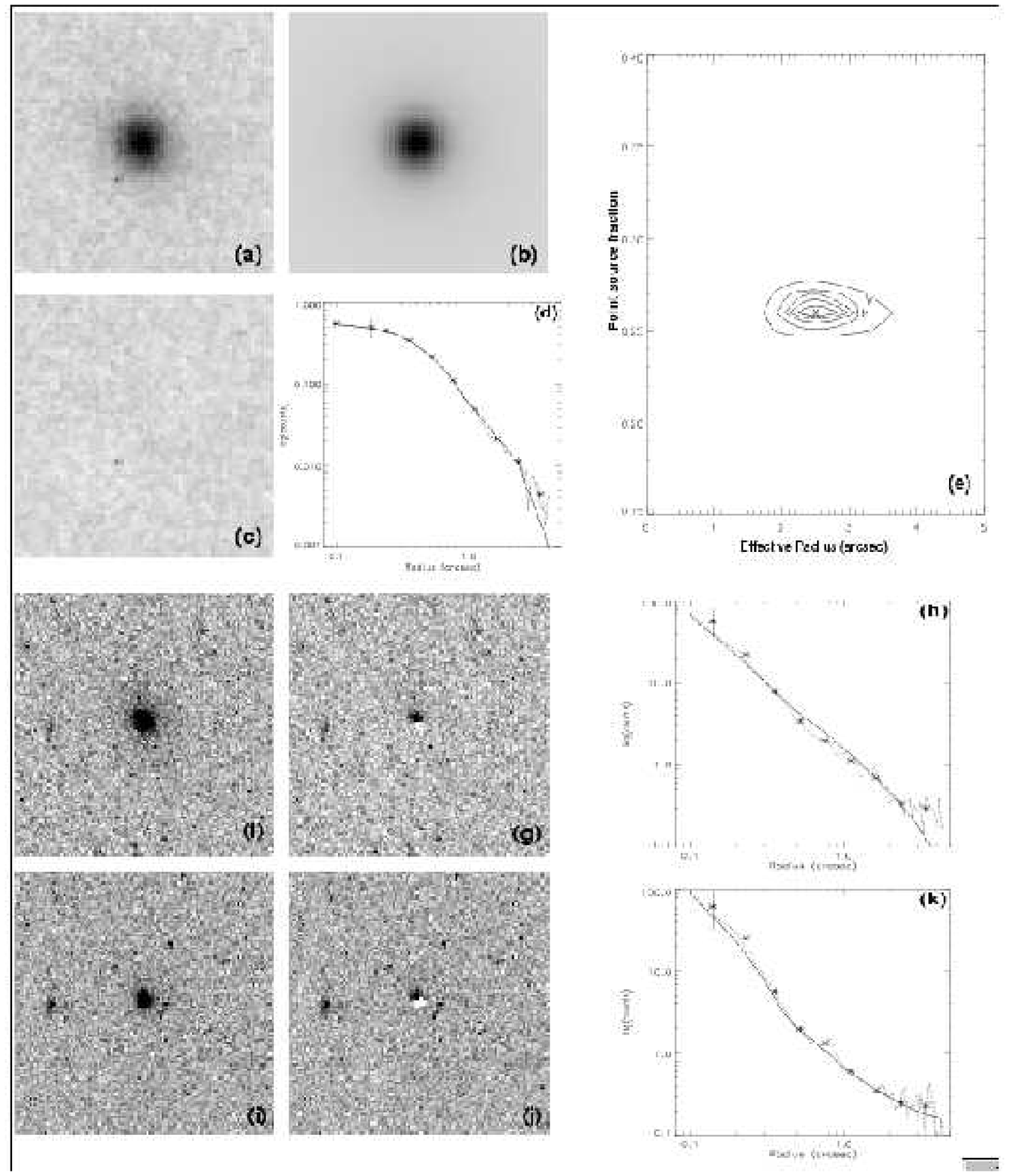}
\end{center}
\caption{Host galaxy morphology fits for the galaxy 6C1217+36. The $K-$band
  galaxy image is presented in frame (a), with the model galaxy given
  in frame (b) and the residuals after subtracting the best fitting model from
  the data displayed in frame (c).  Frame (d) illustrates a 1-d cut
  through the data (dotted line and binned points) with the best-fit model (solid line)
  overlaid. Frame (e) displays the minimum reduced $\chi^2$ (marked with a
cross), plus contours for $\chi^2_{min}
+ 1$, $\chi^2_{min} + 2$, $\chi^2_{min} + 3$, $\chi^2_{min} + 5$ \&
$\chi^2_{min} + 10$. For models with a point source fraction of 0.0\% (very
unlikely for this particular source),
the minimum value of $\chi^2$ lies at $r_{\rm eff} \sim 0.55$\arcsec.
Frame (f) displays the HST data for this source in the F814W filter.  The residual flux
  after subtraction of the best fit model galaxy is displayed in frame
  (g), and frame (h) illustrates a 1-d profile of these data (dotted
  line and binned points) with the best fit model (solid line).
The same data for the F606W filter are displayed in frames (i), (j) \& (k).
\label{Fig: 8}}
\end{figure*}
\begin{figure*}
\vspace{6.2 in}
\begin{center}
\includegraphics{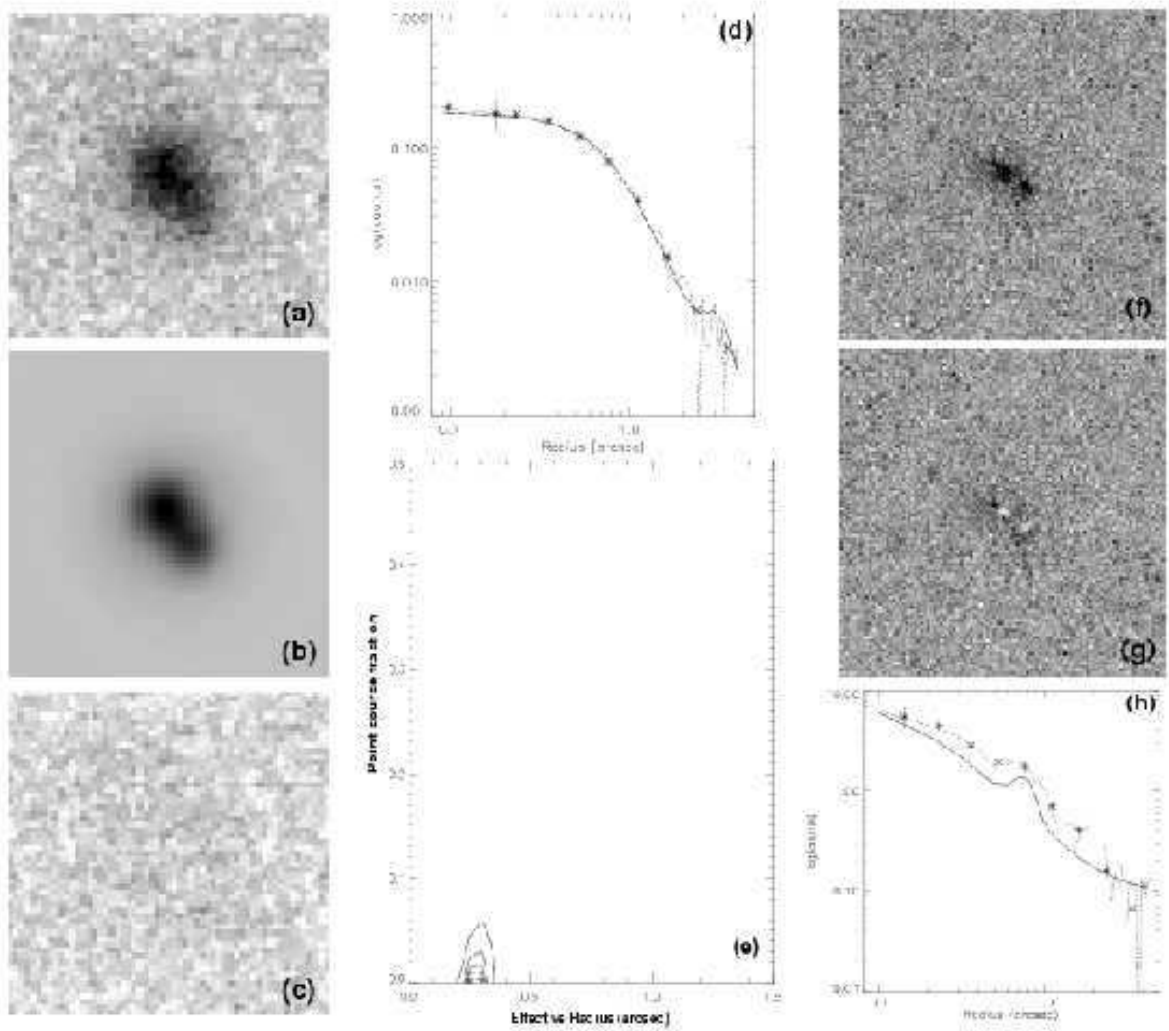}
\end{center}
\caption{Host galaxy morphology fits for the galaxy 6C1204+35.  The fitting of this source
  also included the presence of an unresolved companion object, the
  flux and position of which were additional free parameters. The $K-$band
  galaxy image is presented in frame (a), with the model galaxy given
  in frame (b) and the residuals after subtracting the best fitting model from
  the data displayed in frame (c).  Frame (d) illustrates a 1-d cut
  through the data (dotted line and binned points) with the best-fit model (solid line)
  overlaid. Frame (e) displays the minimum reduced $\chi^2$ (marked with a
cross), plus contours for $\chi^2_{min}
+ 1$, $\chi^2_{min} + 2$, $\chi^2_{min} + 3$, $\chi^2_{min} + 5$ \&
$\chi^2_{min} + 10$. 
  Frame (f) displays the HST data for this source.  The residual flux
  after subtraction of the best fit model galaxy is displayed in frame
  (g), and frame (h) illustrates a 1-d profile of these data (dotted
  line and binned points) with the best fit model (solid line). 
\label{Fig: 9}}
\end{figure*}

\begin{figure*}
\vspace{7.85 in}
\begin{center}
\includegraphics{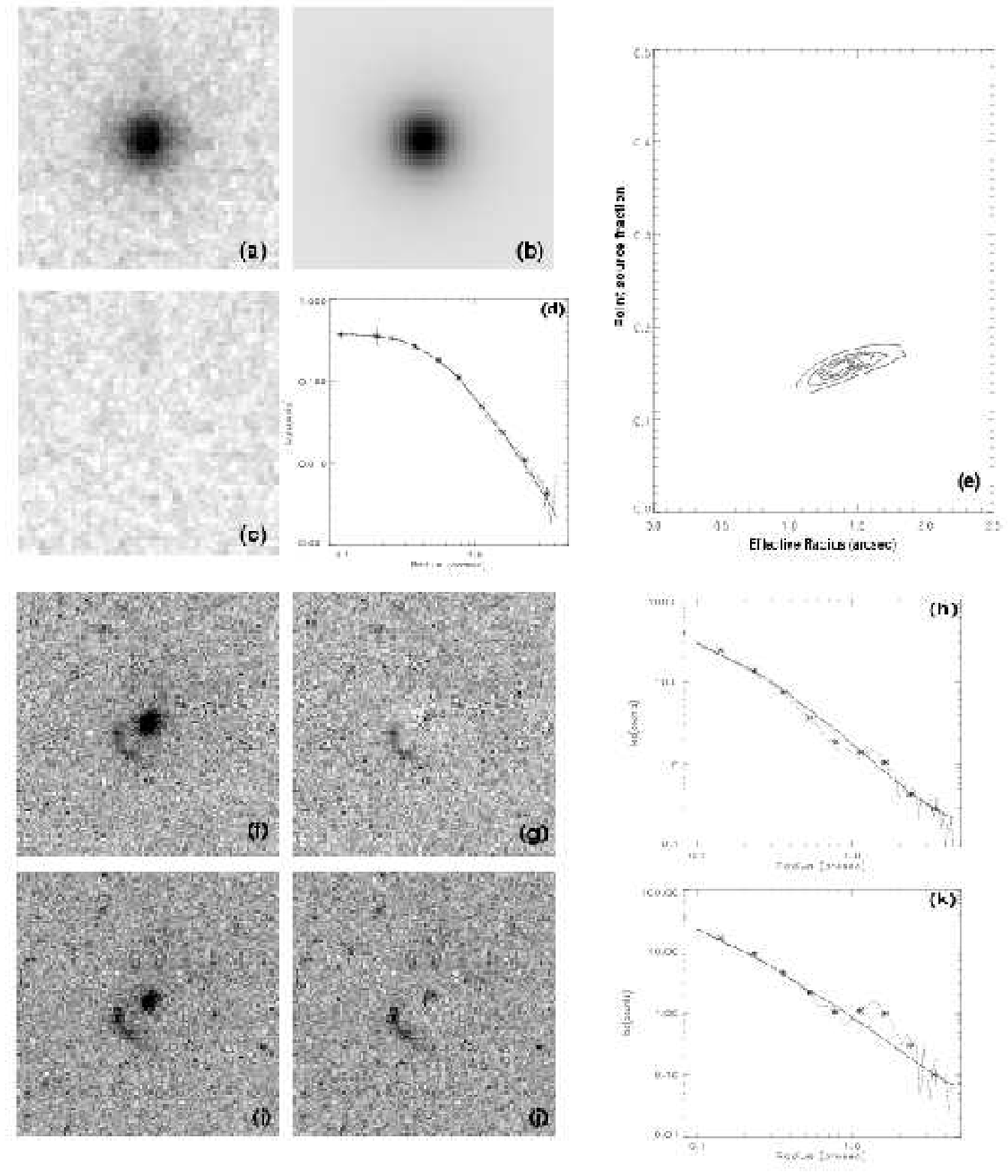}
\end{center}
\caption{Host galaxy morphology fits for the galaxy 6C1257+36. The $K-$band
  galaxy image is presented in frame (a), with the model galaxy given
  in frame (b) and the residuals after subtracting the best fitting model from
  the data displayed in frame (c).  Frame (d) illustrates a 1-d cut
  through the data (dotted line and binned points) with the best-fit model (solid line)
  overlaid. Frame (e) displays the minimum reduced $\chi^2$ (marked with a
cross), plus contours for $\chi^2_{min}
+ 1$, $\chi^2_{min} + 2$, $\chi^2_{min} + 3$, $\chi^2_{min} + 5$ \&
$\chi^2_{min} + 10$.  For models with a point source fraction of 0.0\%,
the minimum value of $\chi^2$ lies at $r_{\rm eff} \sim 0.55$\arcsec.
Frame (f) displays the HST data for this source in the F814W filter.  The residual flux
  after subtraction of the best fit model galaxy is displayed in frame
  (g), and frame (h) illustrates a 1-d profile of these data (dotted
  line and binned points) with the best fit model (solid line).
The same data for the F606W filter are displayed in frames (i), (j) \& (k).
\label{Fig: 10}}
\end{figure*}

We have repeated the fitting
process for our WFPC2 HST observations of a number of 6C sources (see
Paper I for details), excluding
those with the most complicated UV morphologies (6C1129+37) and lowest
signal to noise (6C0825+34).  Point spread functions were determined
using two methods: firstly by extracting the profiles of unsaturated
stellar objects on the WF3 fields, and secondly via the \textsc{tinytim}
software.  No significant differences were
observed between the data PSFs and the \textsc{tinytim}\, 
 PSFs; the latter were used for this analysis.

For the remaining galaxies, we have masked out the regions of the
image which contain the aligned emission. However, this is only
practical for emission which is well separated from the host
galaxy. Any remaining aligned emission, particularly within the
central few arcsec, may potentially skew the fit.   For most sources
we do obtain good 2-d fits.  But for some others, in spite the high
resolution of the HST data, accurate fitting becomes impossible; this
is generally due to the presence of either dust lanes, bright
aligned/excess UV emission coincident with the host galaxy, or the low
signal to noise of the underlying galaxy.  In these cases the
$K$-band fits are used to fix the characteristic radius; this allows
the best-fit nuclear point source contribution to be determined, and
also permits a clear examination of the excess UV emission. 
The resulting models/residuals produced by this
analysis are displayed in figures 1-10, and the 
overall fit parameters are given in table 1.  

One potential pitfall of this approach is that the observed effective
radii of such galaxies may vary with wavelength. The host galaxies of powerful radio
sources typically display bluer colours than those of other elliptical
galaxies; in addition to the obvious effects of a nuclear point source
or the alignment effect, excess blue emission may be spatially extended throughout
the galaxy (e.g. Smith \& Heckman 1989; Govoni et al 2000). There is
also some evidence that powerful radio galaxies become bluer towards
their centres, contrary to the behaviour of normal elliptical galaxies
(Mahabal, Kembhavi \& McCarthy 1999).
If this is the case, the effective radii of the galaxies modelled in
the rest frame UV would be smaller than those measured using our longer
wavelength $K$-band observations,
and the resulting nuclear point source contributions could have
been overestimated.


\subsection{Notes on individual sources}

\subsubsection*{6C0825+34}
6C0825+34, at a redshift of z=1.47, is the most distant source
of the sample.  The signal--to--noise level for this source is the
lowest in the sample, and fitting a simple de Vaucouleurs profile
suggests an effective radius of 0.2$\pm$0.05\arcsec. However, if we also
allow the inclusion of a nuclear point source contribution, neither the
effective radius nor the point source fraction can be accurately
constrained. The rest-frame UV image of this galaxy (which shows a small,
faint galaxy with a secondary emission peak aligned with the radio
source axis; Inskip et al 2003) does not provide much help in
accurately constraining the structural parameters of this object.

\subsubsection*{6C0943+39}
Fitting this galaxy solely with a simple de Vaucouleurs elliptical
galaxy profile results in a best fit effective radius of 0.15\arcsec,
which is in 
good agreement with the value of $r_{\rm eff}=0.27^{+0.33}_{-0.14}$\arcsec
given by the previous fitting of RER98.
However, the data greatly favour a fit including a nuclear point
source contribution; the best fit values being an effective radius of
$1.9\pm0.4$\arcsec and a point source percentage of 41.5\%.  Profile
fitting of the WFPC2 observations of this source is not possible,
due to the extensive aligned emission surrounding this source.
However, fitting using the best-fit value of $r_{\rm eff}$
already obtained from our IR data suggests a point source contribution
in the F702W filter of 36\%; the subtraction of this model galaxy \&
point source does not leave any obvious residual 
features other than the aligned emission already apparent in the
images (Table 1; Fig.~\ref{Fig: 1}).

\subsubsection*{6C1011+36}
Once again, although the fit obtained with a pure de Vaucouleurs
profile ($r_{\rm eff} = 0.25$\arcsec) was in good agreement with the
RER98 result ($r_{\rm eff} = 0.19^{+ 0.09}_{- 0.11}$\arcsec), the best
fit 
for 6C1011+36 also consists of a combination of point 
source (28.8\%) and elliptical galaxy ($r_{\rm eff} = 1.0^{\prime\prime}$). 
The aligned emission surrounding this source is particularly luminous
and difficult to completely mask, rendering a fit of the rest frame UV
morphology impractical. Using the $K$-band effective radius
suggests a point source contribution of 25\%. After subtraction of the
model galaxy, both the luminous aligned emission to the north west of
the galaxy and fainter emission to the south are
clearly visible in the residual image (Fig.~\ref{Fig: 2}).

\subsubsection*{6C1017+37}
This galaxy was relatively easy to fit, and our resulting value of
$r_{\rm eff}=0.2$\arcsec is again in close agreement with the RER98
value ($r_{\rm eff} = 0.12^{+ 0.19}_{- 0.06}$\arcsec).
We have also carried out profile fitting of the WFPC2 observations of
this source, masking out the aligned emission as well as possible.
The resulting fit is in good agreement with that of our IR data,
giving an effective radius of 0.28\arcsec and a point source
contribution of 12\%.  However, we believe this larger size may
reflect 
incomplete masking of aligned emission very close to the host galaxy.
Fixing the effective radius at $r_{\rm eff}=0.2$\arcsec suggests a
slightly lower point source contribution of 9\%; the full structure of
the aligned emission is clearly visible in the residual image
(Fig.~\ref{Fig: 3}).

\subsubsection*{6C1019+39}
In addition to lying at the lowest redshift in the sample
($z=0.922$), the UKIRT and HST images of 6C1019+39 show it to have the
most noticeable ellipticity of the galaxies in this sample. Fitting
the $K$-band data clearly rules out any significant point
source contribution, but also suggests a rather large ellipticity of
0.69, and an effective radius of 1.18\arcsec along the major
axis. (However, we see little change if we fit over
a range of $\epsilon$: the effective radius converges to
a minimum value of 1.0\arcsec as $\epsilon \rightarrow 0$ and the model
galaxy becomes spherical.)  Similar results are obtained for all three
parameters using the HST data, which suggest effective radii of
1.16\arcsec and 1.04\arcsec for the F606W and F814W filters 
respectively.  The residual image also allows us to see several faint
clumps of excess UV emission surrounding this galaxy; although this source
displays very little aligned emission, it does possess very luminous line
emission likely triggered by a combination of both AGN photoionization
and shocks associated with the radio source (Inskip et al 2002b).

\subsubsection*{6C1100+35}
Along with 6C0825+34 and 6C1204+35, this source is one of the three
higher redshift sources in the sample.  Due to the unusually strong
radio emission along the jet axis on one side of this source, it is
plausible that this source may be orientated closer to the line of
sight than the average for this sample, and a high point source
contribution would therefore not be unexpected. Due to the close
proximity of a second object, we fit both objects together.  The
resulting best fit suggests a galaxy with $0.16$\arcsec with a point
source contribution of $15 \pm 14$\%.

As this source displays no obvious aligned emission, it is also useful
to repeat the fitting process for the HST data. However given the
strong point source contribution to 6C1100+35 in the rest-frame UV and
the relatively small sizes of both sources, we unfortunately find
that the effective radius and point source contribution become
considerably degenerate.  Using the best fitting results obtained from
our $K-$band observations gives a maximum point source contribution in
the F814W filter of 41\%.

\subsubsection*{6C1129+37}

Our $K-$band image of this source clearly reveal a pair of elliptical
galaxies, of roughly similar luminosity.  As was the case for
6C1100+35, these two galaxies need to be fitted simultaneously, giving
a best fit value of 1.8\arcsec for the radio galaxy and a 
point source contribution of 18\%.
Due to the extreme aligned emission surrounding this source in the
rest frame UV, accurate fitting of our HST data is not possible. We
have however attempted the removal of the host galaxy assuming the
$K$-band effective radius and zero point source, to permit examination
of the remaining aligned emission (Fig.~6g).

\subsubsection*{6C1204+35}

For this source, we find best fit values of $r_{\rm eff}=0.85$\arcsec
and a point 
source fraction of 13\% (our results for a pure de Vaucouleurs profile
with zero point source are again in good agreement with the RER98 results, 0.45\arcsec cf. $0.51^{+0.10}_{-0.14}$\arcsec).
The WFPC2 observations of 6C1204+35 cannot be easily
fitted.  Fig.~\ref{Fig: 7} displays the residuals produced after fitting
this galaxy with the best-fit effective radius found for the IR data.
These clearly show that there is either a dust lane/disk lying across the
galaxy perpendicular to the radio axis, or (perhaps less likely)
bright aligned emission 
lying along the radio axis through the centre of the galaxy.
Interestingly, S\'{e}rsic profile fitting of
the $K$ band observations of this system imply values of $n \sim 2$,
$r_{\rm eff}=0.73$\arcsec and a higher point source contribution of
22\%. This suggests that although the nuclear emission is strong, the
emission in the inner ($r < r_{\rm eff}$) regions of the host galaxy
has a shallower profile than a de Vaucouleurs elliptical, perhaps due
in part to dust extinction.

\subsubsection*{6C1217+36}
Once again, although our pure de Vaucouleurs profile fit
($r_{\rm eff}=0.55$\arcsec) is in good agreement with the RER98 fit
($0.44^{+0.18}_{-0.20}$\arcsec), the data very much favour an
additional point source contribution, giving best fit values of
$r_{\rm eff}=2.5$\arcsec and 25\%.   S\'{e}rsic profile fitting of
this source predicts $n = 5.4$, a similar point source contribution of
22\% and a somewhat larger $r_{\rm eff}=3.9$\arcsec.
 We also expect a high point source
contribution in our HST images, based on the very blue colours of this
source coupled with the lack of any large scale aligned emission.
Point source contribution and effective radius again prove difficult
to disentangle from our HST images.  Given the excess emission visible
in the residual image (offset slightly to the north from the centre of
the galaxy), and the low signal to noise of the underlying host galaxy, this is
hardly surprising.

\subsubsection*{6C1256+36}
This source displays the most elongated $K$-band morphology; the major
question for this source is whether this elongation is due to a strong
$K$-band alignment effect (such as that seen for some sources in the
more powerful 3CR subsample of Best et al 1997), the natural elongation of the elliptical
host galaxy, a close pair of galaxies, or a single galaxy combined
with an unresolved point source.  Single object fits could not provide
an adequate fit to the data, but much better fits were obtained
fitting two objects simultaneously. The positions of the two
components in the $K$-band are coincident with the two peaks
visible in our HST image. Whilst fitting this source, the
positions of both objects were free parameters, as were their
effective radii and fractional point source contributions.  The best
fits were obtained for an unresolved ``companion'' object, hardly
surprising given its point-like appearance on the HST image. Overall,
this stellar object has a flux equivalent to roughly 34\% of the total
$K$-band flux from this galaxy (i.e. a 25\% contribution to the
aperture flux for this object).  Whilst the point-like nature of the
adjacent object (hereafter assumed to be a foreground star) is clear from the
WFPC2 observations of this source, the host galaxy itself is
relatively faint and is not well suited to profile fitting.  Fitting
using the best-fit parameters of the IR imaging produces negligible
residuals; in this case the flux of the adjacent object accounts for
17\% of the total flux in the aperture (i.e. equivalent to 20\% of the flux from the
galaxy itself).

\subsubsection*{6C1257+36}

As is the case for the majority of our objects the best fit for this
object consists of a combination of elliptical galaxy
($r_{\rm eff}=1.39$\arcsec) and nuclear point source (16\%); if
we exclude the point
source contribution (which clearly gives a much worse fit than the
models including 
a point source) the suggested effective radius (0.55\arcsec) is
similar to (although somewhat higher than) the
RER98 value ($0.37^{+0.07}_{-0.08}$\arcsec).  Fitting this system with
a S\'{e}rsic profile suggests values of $n = 6.1$, $r_{\rm
  eff}=1.8$\arcsec and a point source contribution of 12\%.
The presence of
excess UV emission to the north of the galaxy (as well as to the east,
as can be seen in the residual images: Fig.~\ref{Fig: 10})
makes fitting the HST images less straightforward; with the exception
of these features, the effective
radius implied by the $K-$band fitting seems to provide a reasonable
fit, suggesting point source contributions of 11\% in the F606W
filter, and 3.5\% in the F814W filter.

\subsection{Comparison with other radio source samples}

Analysis of the radio galaxy $K-z$ relation (Inskip et al 2002a) shows
that whilst the 6C and 3CR galaxy populations are indistinguishable at
low redshifts, the two populations differ in their mean magnitude by
approximately $K \sim 0.6$mag at higher redshifts.   If galaxy mass is
the sole cause of this difference between the samples, the more
powerful radio sources would be expected to be hosted by  $\sim 75$\%
larger elliptical galaxies.  Comparison of the results of RER98 (mean
effective radius of 3.6kpc for elliptical host galaxies, excluding the
double source 6C1129+37) with those of BLR98 for the 3CR galaxies
(14.7kpc) suggests a much larger difference than this, but the
inclusion of the presence of nuclear point sources in our galaxy
fitting procedure can greatly improve on this result. The mean
effective radius
we derive for the $z \sim 1$ 6C galaxies is $\sim 10.5$kpc, which is
less than that of the 3CR data by a little under 50\%. 
This corresponds to an expected $K$-band magnitude difference of $\sim
0.5$mag, in excellent agreement with the observed $K-z$ relation.

\begin{table*}
\caption{Typical galaxy sizes for a number of different radio source
samples.  Columns 1-3 provide details of each sample: a reference,
the radio source sample from which the data were selected and the mean redshift. Column 4 gives the median
effective radius, and column 5 the mean effective radius and
standard error on the mean. The average percentage nuclear point
source contribution, where available, is listed in column 6.  All data have been converted to the same
cosmological model of $\Omega_0=0.3$,$\Omega_\Lambda=0.7$ and
$H_{0}=65\,\rm{km\,s^{-1}\,Mpc^{-1}}$ where necessary. }
\begin{center} 
\begin{tabular} {lcccr@{$\pm$}lc}
Reference & Sample & Redshift & Median $r_{\rm eff}$ (kpc)&
\multicolumn{2}{c}{Mean  $r_{\rm eff}$ (kpc)} & Point source percentage\\ \hline
Inskip et al 2005 & 6C & 1.11 & 9.35 & 10.45 & 1.94 & $16.1 \pm 4.0$\% \\
Pentericci et al 2001 & MRC & 2.06 & 5.85 & 5.96 & 1.77 & --\\
Zirm et al 2003 & 3C & 1.16 & 8.19 & 8.85 & 1.35& $1.2 \pm 0.5$\% \\
BLR98  & 3C & 0.93 & 15.1 & 14.72 & 1.20& $7.3 \pm 2.7$\% \\
McLure \& Dunlop 2000 & 3C & 0.89 & 9.45 & 11.42 & 1.55 &--\\
McLure et al 2004 & 3C, 6C, 7C, TOOT & 0.50 & 13.89 & 14.97 & 0.98&--\\
McLure et al 2004 & 3C & 0.49 & 17.88 & 19.41 & 1.62&--\\
McLure et al 2004 & 6C & 0.49 & 13.25 & 13.58 & 1.53&--\\
McLure et al 2004 & 7C & 0.53 & 11.15 & 11.21 & 1.27&--\\
McLure et al 2004 & TOOT & 0.48 & 10.98 & 12.53 & 1.67&--\\
Roche \& Eales 2000 & 3C (all sources) & 0.20 & 9.56 & 15.72 &
4.08 &--\\
  & 3C (disks excluded) & 0.20 & 10.56 & 17.34 & 4.55 &--\\
McLure et al 1999 & RGs, RLQs, RQQs$^1$ & 0.19 & 7.42 & 9.51 & 1.03&--\\
McLure et al 1999 & RGs only$^1$ & 0.19 & 12.66 & 12.58 & 2.27&--\\
Zirm et al 2003 & 3C & 0.15 & 4.48 & 5.49 & 1.11 &--\\\hline

\multicolumn{6}{l}{Notes:}\\
\multicolumn{6}{l}{[1] Full sample consists of a mixture of radio galaxies,
radio loud quasars and radio quiet quasars.}\\
\vspace{-5pt}\\
\end{tabular}
\end{center}
\end{table*}

However, although these results would indicate a difference in mean
radii between the two samples, Kolmogorov-Smirnov tests suggest that
the distribution of radial sizes for the two samples are in
fact not significantly different.
In addition, alternative estimates of the mean radii for 3CR sources
have suggested lower values than those of BLR98: McLure \& Dunlop
(2000) derived a mean radius of 11.4kpc from WFPC2 observations of a
subsample of the 3CR galaxies used by BLR98, and Zirm et~al (2003)
used NICMOS observations of 3CR galaxies to derive a mean radius of
8.85kpc. These results, together with the lack of significance in the
Kolmogorov-Smirnov test, suggest that the radii of the 3CR and 6C
galaxies are very similar.
The range of results obtained is not unsurprising given the small
sizes of the samples which have been studied, and the observed
large scatter within the two galaxy samples from which these data were
drawn.

Our observed similarity in host galaxy properties is confirmed by other
recent work on the wider radio galaxy population. A summary of the key results
from a number of different studies of radio galaxy samples at different
redshifts and radio powers has been collated in Table 2.   Although not
statistically significant, the average effective radii found 
for these data are typically somewhat larger than those of our 6C 
subsample.
Continuing the comparison with lower redshift 3CR galaxies,
matched in radio power to the $z \sim 1$ 6C subsample (e.g. McLure et
al (1999); Roche \& Eales (2000)), the emerging picture is that a
radio source similar to our $z \sim 1$ 6C sample in power will be hosted by a $\sim 10$kpc
elliptical galaxy regardless of the cosmic epoch at which it exists.
Although galaxy mass may play some role in
explaining the drop in $K$-band luminosity between the 6C/3CR samples,
it seems that it cannot account for it completely.

\begin{figure}
\vspace{1.9 in}
\begin{center}
\includegraphics{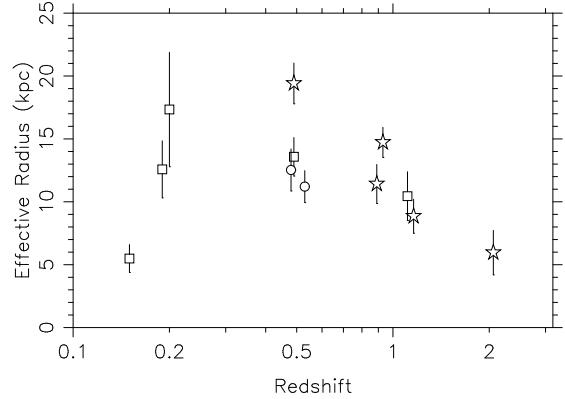}
\end{center}
\caption{Mean effective radius vs. redshift for the radio galaxy
samples of Table 2.  The symbols denote the typical radio
  luminosities of the samples: the stars have $L_{151} < 10^{26}
  \rm{W} \rm{Hz}^{-1} \rm{sr}^{-1}$, squares have $10^{26} < L_{151} <
  10^{27} \rm{W} \rm{Hz}^{-1} \rm{sr}^{-1}$, and circles have $L_{151} > 10^{27}
  \rm{W} \rm{Hz}^{-1} \rm{sr}^{-1}$.  Error bars are the standard error on
  the mean for each sample.
\label{Fig: grow}}
\end{figure}

Despite this, there is still some evidence in favour of a mass--radio
luminosity relationship. McLure et al (2004) provide an analysis of
four radio galaxy samples of different powers 
at $z \sim 0.5$.  For these data, the 3CR sources were found to
have larger sizes (median 17.9kpc) than the 6C sources
(13.3kpc), which were in turn more massive than the still less
powerful 7CRS (11.2kpc) and TOOT (11.0kpc) radio
sources, suggesting a weak correlation between radio luminosity and
bluge/black hole mass within their $z \sim 0.5$ sample.  K--S tests on
these data confirm that the 3CR sources are indeed larger on average
than the 6C/7CRS/TOOT sources at a signifcance level of $> 95$\%. 
There is no significant size difference between the 3CR and 6C data
alone, or between the 6C sources and the lower radio power 7CRS/TOOT
sources, emphasising the weakness of this trend and the ease with
which it can be swamped by the relatively large scatter in galaxy
properties within each sample.  The 6C galaxies at $z \sim 0.5$ appear
to be larger on average than those at $z \sim 1$, which if truly
representative of evolution in the galaxy properties, could indicate
continued growth (e.g. via mergers) of the host galaxy population at
$z < 1$.   However, the $z \sim 0.2$ sources appear comparable in size
to those at $z \sim 1$, which suggests that the larger sizes of the $z
\sim 0.5$ sources may simply be anomalous.  We have also plotted the
data presented in Table 2 as a function of redshift (Fig.~\ref{Fig:
  grow}).  The scatter in the data is large, and we have not
included any systematic errors arising from the different methods and
quality of datasets used. However, this plot does provide a useful
illustration of an apparent gradual growth of the host galaxy population,
from $z \sim 2$ up to at least $z \sim 1$, perhaps continuing to later times.

A final interesting point of note is that given that any difference
between the 6C and 3CR host galaxy masses is 
far less than their difference in  radio power (a factor of $\sim 5$
at $z \sim 1$), it might be expected that the contribution from a
nuclear point source would on average be lower for the 6C sources;
this assumes that any unresolved nuclear emission associated with the
AGN scales with radio power.  We find mean point source contributions of
$16\pm4$\% and $7\pm3$ for the 6C and 3CR samples respectively.
Surprisingly, Kolmogorov-Smirnov tests show that the greater relative
point source contribution for the 6C data is in fact weakly
significant at the 90\% level, although due to the small sample size
this result could be dominated by one or two objects.  Interestingly, if we
compare the core--lobe ratio of radio flux between the 6C and 3CR
subsamples (Best {\it et al} 1997b, Best {\it et al} 1999), the values
obtained are very different.  The 6C sources have an average core flux
fraction of 0.026 (excluding the very high core fraction of the
unusual radio source 6C1217+36), whereas the mean 3CR galaxy core
fraction is 0.0045.  This could imply that the 6C sources are observed
at an angle closer to the line of sight than the 3CR sources, and the
AGN are less heavily obscured, thereby increasing the point source
contributions to the observed 6C magnitudes.

The unresolved point source emission observed in many of the galaxies
in our sample may be due to several processes: the AGN itself, nuclear emission
lines or an associated unresolved nuclear starburst.  We have compared the
nuclear point source contributions in the 6C and 3CR $z \sim 1$
samples with the emission line luminosities of our spectroscopic observations (Inskip et
al 2002b,c; Best et al 2000a,b), and find no correlation.  This
suggests that nuclear emission lines are unlikely to 
be responsible for the observed point source emission. Our data do not
allow us to distinguish between the other two possibilities of AGN or
nuclear starburst.
The impact of this nuclear point source emission on the
galaxy colours will be considered further in paper III of this series.

\section{The alignment effect in the 6C subsample}
\subsection{Distributions of the excess UV emission}      

The aligned structures surrounding 6C and 3CR radio galaxies at $z
\sim 1$ display a wide variety of different features.  The most
luminous features observed are bright knots of continuum emission;
these are seen near several galaxies (e.g. 6C1011+36, 6C1129+37,
3C356, 3C368 and others). Another type of prominent feature often
observed is a bright arc-like structure, extending from the host
galaxy, and occasionally linked to other bright features.  A
particularly good example of such an arc can be seen in images of
3C280; arcs are also observed in the HST images of 6C1017+37 and
6C1204+35.  A third 6C radio source, 6C1257+36, displays a small, very
blue arc--like structure, apparently associated with a bright knot in
the radio emission (Paper I).  Fainter linear structures, or filaments of
diffuse emission can be seen near, for example, 6C0943+39 and 3C352.
Other commonly observed features are faint linear extensions or tails
projecting out from the host galaxy, such as those observed in images
of 6C0825+34 and 3C22. These are not however due to optical emission
from a radio jet.  In general, some amount of faint diffuse emission
is observed around the majority of sources.  Usually, each source
exhibits one or more of these distinct features. Whilst some sources 
appear completely passive (e.g. 6C1100+35, 6C1217+36 and 3C65),
after subtraction of the best-fit host galaxy model we often
see signs of excess emission towards or offset from the centre of the
host galaxy in the residual image (e.g. 6C1217+36 in particular).
 
In addition to the aligned emission observed around these sources, the 
appearance of several galaxies suggests that the UV emission from
these sources has been affected by the presence of dust.  Extensive
dust lanes lying perpendicular to the radio axis, such as those
observed in UV imaging observations of low redshift radio galaxies
(Allen et al 2002; de Koff et al 1996), could plausibly explain certain
morphological features of the host galaxies of high redshift radio sources
such as 3C252 \& 3C324 (Best et al 1997; Longair, Best \&
R\"{o}ttgering 1995), 6C1204+35, and at still lower radio powers,
7C1748+6731 (Lacy et al 1999). It is worth noting that although the presence of a dusk
disk could create the false appearance of a small scale alignment
effect, this is only likely to be the case for a single object in our
sample, 6C1204+35. 

The final feature
revealed by the imaging observations is the presence of other nearby galaxies
which may lie at the same redshift.
Four of the eleven 6C sources have companion galaxies at a
close projected distance ($< 20$kpc; {\it see Paper I in this series}); a few of
these sources clearly seem likely to be interacting with their companions.
Similarly, a number of the $z \sim 1$ 3CR sources appear to have
nearby companions (Best 1996); in several cases companion galaxies
show evidence (e.g. increased star formation) for having been 
perturbed by the growing radio source.  This issue will be discussed
in more depth in Paper III of this series.

Whilst the infrared emission of the $z \sim 1$ 3CR host galaxies has
been found to show a weak alignment effect (Zirm, Dickinson \&
Dey 2003; BLR98 and references therein), we do not see any noticeable
alignment in the $K$-band for our 6C subsample (nor is one seen in the
case of low redshift 3CR sources (de Koff et al 1996; 
Martel et al 1999), for which the alignment effect is
weaker than at high redshifts at all wavelengths).  
The percentage contribution of the IR aligned component at high
redshifts is
known to be relatively low in any case: BLR98 and Rigler et al (1992)
find typical contributions of 11\% and 10\% of the total IR flux
respectively.  Although this in itself is insufficient to account for the
$K$-band magnitude difference between the $z \sim 1$ 6C and 3CR
samples, the infrared aligned component is still notable by its
absence in the 6C subsample. 

Having modelled the galaxy morphologies, the simplest method of
quantifying the aligned emission is to evaluate the residual flux
within some aperture.  For the sources for which we have adequate
galaxy models in the rest frame UV, the resulting residual flux due to
excess UV emission within a 4\arcsec diameter aperture has been
determined, and is tabulated in Table 3.  

\begin{table}
\caption{Flux in residual images after subtraction of model
galaxies. Results are given as a percentage of the total 4\arcsec
diameter flux (column 3) and as a magnitude correction between the total
flux in the aperture and the flux from the model galaxy alone (column
4). The data for 6C1129+37 are presented in italics, as the accuarcy
of the host galaxy subtraction is unclear for this source.}
\begin{center} 
\begin{tabular} {lccc}
Source & Filter &  Residual Flux & Magnitude Correction \\\hline
\it{6C0825+34} & \it{F814W} & --   & -- \\
6C0943+39 & F702W & $39 \pm 4$\% & $0.55 \pm 0.04$ \\
6C1011+17 & F702W & $20 \pm 3$\% & $0.24 \pm 0.03$ \\
6C1017+37 & F702W & $22 \pm 2$\% & $0.26 \pm 0.02$ \\
6C1019+39 & F814W & $6 \pm 1$\%  & $0.07 \pm 0.01$ \\
          & F606W & $4 \pm 2$\%  & $0.05 \pm 0.02$ \\
6C1100+35 & F814W & $4 \pm 2$\%  & $0.04 \pm 0.02$ \\
\it{6C1129+37} & \it{F702W} & \it{63 $\pm$ 4\%} & \it{1.09 $\pm$ 0.05} \\
6C1204+35 & F814W & $28 \pm 3$\% & $0.36 \pm 0.03$ \\
6C1217+36 & F814W & $10 \pm 3$\% & $0.11 \pm 0.02$ \\
          & F606W & $9 \pm 2$\%  & $0.10 \pm 0.03$ \\
6C1256+36 & F702W & $43 \pm 6$\% & $0.62 \pm 0.06$ \\
6C1257+36 & F814W & $5 \pm 2$\%  & $0.05 \pm 0.02$ \\
          & F606W & $36 \pm 4$\% & $0.49 \pm 0.04$ \\
\vspace{-5pt}\\
\end{tabular}
\end{center}
\end{table}

\subsection{Quantifying the Alignment Effect}

A very basic analysis of galaxy colours in different filters can be
useful for quantifying the total excess UV 
emission, but it is also of crucial importance that the spatial
distributions of the extended emission regions are understood. In
particular, the total size of the extended structures and the degree
to which they are aligned with the radio source axis vary greatly
within both the 6C and 3CR subsamples.
The strength of the alignment effect can be quantified using
measurements of these different features, using the method of BLR98.  This 
utilises the difference in position angles of the radio axis and the
aligned emission, and the total elongation of the extended
structures.  
 
The {\it alignment strength} is defined as:
\begin{equation}
a_s = \epsilon \left( 1 - \frac{\Delta \theta}{45} \right).
\end{equation}
$\epsilon$ is the ellipticity of the rest frame UV emission on the HST images; this is
defined as $\epsilon = [1 - q^2/p^2]^{1/2}$, where $p$ and $q$ are the
lengths of the semi-major and semi-minor axes respectively of the
ellipse which provides the best-fit to the large scale UV excess.  $\Delta
\theta$ is the difference in the position angles of the radio axis and
the semi-major axis of the rest-frame UV emission.  The alignment
strength can therefore have values in the range $-1 \leq a_s \leq 1$.
Sources with 
$\Delta \theta < 45^\circ$ have a positive value for $a_s$.
Sources which are misaligned with the radio axis have $\Delta \theta >
45^\circ$ and $-1 \leq a_s \leq 0$.  For perfectly round galaxies,
$\epsilon = 0$ and $a_s = 0$.  

\begin{table}
\caption{Table of alignment strengths and component alignment
strengths for the 6C subsample. The first two columns list the sources,
and the filters used for the observations. Column 3 gives the measured
number of bright components in the data; where the radio source has a
close companion, we also list (in brackets) the value of $N_c$ after
removal of the companion object.  The ellipticity and position
angle of the fitted ellipses are given in columns 4 and 5
respectively.  Column 6 lists the difference in position angles
between the radio axis and the aligned emission.  These parameters are
then used to determine the alignment strength ($a_s$, column 7) and
the component alignment strength ($a_c$, column 8) for each observation.}
\scriptsize
\begin{center} 
\begin{tabular} {lcccrrr@{.}lr@{.}l}%
Source    & Filter &$N_c$ & $\epsilon$ &\multicolumn{1}{c} {$\theta$}
& \multicolumn{1}{c}{$\Delta \rm{PA}$}    &\multicolumn{2}{c}{$a_s$} & \multicolumn{2}{c}{$a_c$}\\\hline
\vspace{-5pt}\\
\vspace{5pt}
6C0825+34 & F814W  & 1    &   0.47     & -72$^{\circ}$    & 26$^{\circ}$ &   0&198  & 0&198 \\\vspace{2pt}
6C0943+39 & F702W  & 1    &   0.69     & -34$^{\circ}$    & 17$^{\circ}$ &   0&429  & 0&429 \\\vspace{2pt}
6C1011+36 & F702W  & 2    &   0.40     & -44$^{\circ}$    & 28$^{\circ}$ &   0&151  & 0&302 \\\vspace{2pt}
6C1017+37 & F702W  & 1    &   0.32     &  70$^{\circ}$    & 1$^{\circ}$  &   0&313  & 0&313 \\
6C1019+39 & F606W  & 2    &   0.22     & -87$^{\circ}$    & 42$^{\circ}$ &   0&015  & 0&030 \\\vspace{2pt}
          & F814W  & 1    &   0.29     & -76$^{\circ}$    & 53$^{\circ}$ &   -0&052 & -0&052\\\vspace{2pt}
6C1100+35 & F814W  & 1    &   0.10     &  15$^{\circ}$    & 82$^{\circ}$ &   -0&082 & -0&082\\\vspace{2pt}
6C1129+37 & F702W  & 6(6)$^1$    &   0.47     &  56$^{\circ}$    & 18$^{\circ}$ &   0&282  & 1&692 \\\vspace{2pt}
6C1204+35 & F814W  & 1    &   0.57     &   2$^{\circ}$    & 17$^{\circ}$ &   0&355  & 0&355 \\
6C1217+36 & F606W  & 1    &   0.12     &  50$^{\circ}$    & 9$^{\circ}$  &   0&096  & 0&096 \\\vspace{2pt}
          & F814W  & 1    &   0.20     &  30$^{\circ}$    & 11$^{\circ}$ &   0&151  & 0&151 \\\vspace{2pt}
6C1256+36 & F702W  & 2(1)$^2$    &   0.40     &  55$^{\circ}$    & 15$^{\circ}$ &   0&267  & 0&533 \\
6C1257+36 & F606W  & 2    &   0.38     & -32$^{\circ}$    & 16$^{\circ}$ &   0&245  & 0&490 \\\vspace{2pt}
          & F814W  & 1    &   0.39     & -40$^{\circ}$    & 25$^{\circ}$ &  0&173  & 0&173 \\\vspace{2pt}\\
\multicolumn{10}{l}{Notes:} \\
\multicolumn{10}{l}{[1]: The companion galaxy to this source forms only
  part of one of the six discrete components,} \\ 
\multicolumn{10}{l}{and its removal would
  not affect the value of $\rm N_c$ for this source} \\
\multicolumn{10}{l}{[2]:  For 6C1256+36, the removal of the companion
  object would reduce $\rm N_c$ from 2 to 1.} \\
\end{tabular}
\end{center}
\end{table}

In order to determine the orientation and ellipticity of the
emission, ellipses were fitted to the HST images
using the \textsc{iraf} package \textsc{ellipse}. The HST images were first 
smoothed, using a Gaussian with $\sigma = 0.2$\arcsec. 
By smoothing the data, consistent fits to the relative elongation
and alignment of the irregular structures observed around these
sources could be obtained over a large range of different size
ellipses. The
aim of this was to obtain a consistent fit to the galaxy, whilst
including as much of the aligned emission as possible without
introducing too much noise from the sky background.  Without first
smoothing the data, consistent fits would be 
unobtainable, as the presence of a bright peak would skew the results
at a given fitting radius.  This Gaussian width was carefully chosen
to allow accurate fitting of the data without over smoothing.
 The orientation
and ellipticity of the galaxies were taken from the average parameters
of several successive fitted ellipses, with the requirement that these
did not vary significantly, and included the majority of the flux from
the galaxy and any aligned emission.  Once the fitting was completed,
the orientation of the emission was compared with the radio axis, and
a measure of  $\Delta \theta$ obtained.

This process is relatively straightforward and gave clear, consistent
results for the majority of sources.  In the case of 6C1204+35, there
is some doubt as to whether the bright region of emission roughly
4\arcsec\ to the north of the host galaxy is aligned emission, or a
companion galaxy.  Ellipses were fitted to the emission from the
host galaxy without attempting to include this feature.
Full results of our analysis of the
6C subsample aligned structures are presented in Table 4.

\begin{figure}
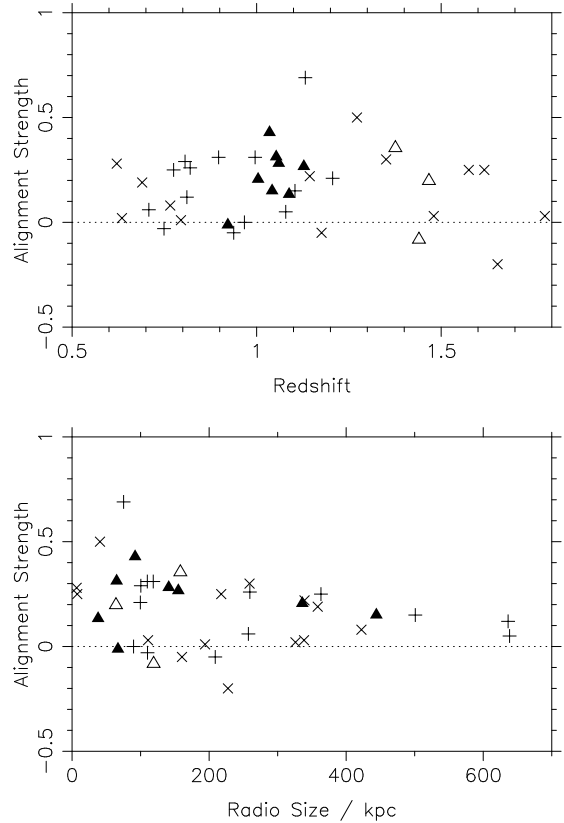

\vspace{4.2 in}
\begin{center}
\includegraphics{as_d.ps}
\includegraphics{as_z.ps}
\end{center}
\caption{Alignment strength vs. redshift (top) and radio size (bottom)
for the 6C and
3CR radio galaxies at $z \sim 1$. 6C sources are represented by the
triangles: filled 
triangles represent the eight sources in the $0.85 < z < 1.2$ spectroscopic
sample (Inskip et al 2002b).  3CR sources are represented by crosses:
the sources in the $0.7 < z < 1.25$ 3CR spectroscopic sample (Best,
R\"{o}ttgering \& Longair 2000a) are marked by `$+$', with the
remaining galaxies marked by `$\times$'.  
\label{Fig: 7_6}}
\end{figure}

The alignment strengths of the 6C galaxies are very similar to those
of 3CR sources at the same redshifts and radio sizes. This can be seen
in Fig.~\ref{Fig: 7_6}, which plots alignment strength vs. redshift
and radio size, for observations made in the filter closest to a
rest--frame wavelength of 3500\AA.  Where observations were made in
two filters which straddle a rest--frame wavelength of 3500\AA, a
weighted interpolation has been used in order to derive the value at
3500\AA.  Both samples seem to show a peak in the alignment strength
at a redshift of just over $z = 1$.  This correlates well with the minimum value
in F707W$-K$ (see Paper III), observed at $z \sim 1.1$; the galaxies displaying the
strongest alignment effect are also bluer in colour due to this excess
UV emission.  The excess UV flux determined in section 3.1 is also
found to be strongly correlated with the alignment strength, at a
significance level of $> 99$\% (excluding 6C1129+37, as the host
galaxy cannot be easily modelled in the rest frame UV). 

In addition to the observed variations with luminosity and redshift,
the alignment strength is also at a maximum for the smaller
radio sources.  Including all
6C and 3CR sources, alignment strength is anti-correlated with radio
size in a Spearman rank correlation test at a significance level of $>
96\%$. This is as expected, given the greater extent of the aligned
structures observed around these smaller radio sources (Best et al 1996;
Paper I; this paper). 
The calculated alignment strengths are based on measurements of the difference in
position angles between the radio and UV emission, and the ellipticity
of the smoothed images.  For the galaxies in both samples in the
redshift range $0.85 < z < 1.5$, the measured ellipticities are
similar at $\epsilon \sim 0.35$. This suggests that the extended
structures are also equally well aligned in both samples. 

Whilst the extended features observed around distant radio sources in
both samples are similar, one major difference is observed.  Extended
strings of bright, knotty emission, well aligned with the radio axis
(such as those observed in 3C266, 3C324 \& 3C368) are less frequently 
seen in the observations of the 6C subsample.   A visual inspection of
the HST images suggests that where these features are observed around
the 6C galaxies, they include fewer discrete components and are
generally located closer to the host galaxy; even 6C1129+37 displays
less extreme features than those of many 3CR galaxies at the same redshift.   

We have therefore quantified some of the morphological
differences between the two subsamples.  Due to the
strong variations observed with redshift in the 3CR subsample, this
analysis has been restricted to the thirteen galaxies in the redshift range of
the 6C subsample, $0.85 < z< 1.5$. Firstly, we consider the number of
bright components in the images of the 6C and 3CR galaxies, and also
the mean separation of these luminous features. 

The number of discrete components of emission was determined using the same definition
as Best (1996),  but with some added restrictions required by the 
fainter 6C galaxies.  The minimum flux level for a bright component
has been set at 10\% of the maximum flux, or at 3$\sigma$ above the
sky noise level, whichever is larger. This prevents sky noise being
identified as a spurious bright component. The \textsc{iraf} package
\textsc{imreplace} was used to blank out pixels below this cut-off
flux level, thus identifying the potential bright components.
Any potential bright
component was then examined using a lower cut-off flux level of 5\% of
the maximum emission.  Confirmation was only given to components which
covered at least 4 adjacent pixels and were unconnected to any other
bright component at this lower flux level.

\begin{table}
\caption{Distribution of aligned emission features. The first half of
this table tabulates the distribution of 
the number of bright emission components for the 6C and 3CR galaxies
in the redshift range $0.85 < z < 1.5$.  Different
values of $N_c$ were obtained in different filters for two sources;
in this analysis we classify 6C1019+39 as having $N_c = 2$, and
6C1257+36 as $N_c = 1$. We do not know whether the unresolved
feature adjacent to 6C1256+36 lies at the same redshift or is a
foreground object, and have classed this object as $N_c=2$ rather
than 1.  The second half compares the
mean component separation for $N_c = 2$ and $N_c \geq 3$. 
 The final section quantifies the range of
physical extents for the aligned emission across the samples,
assigning each galaxies to one of four different size ranges. The
emission around the 3CR
sources is typically more extensive, as expected given their larger
average value of $N_C$.} 
\begin{center} 
\begin{tabular} {cccc} 
&{6C Sample} & {3CR Sample} \\\hline
\multicolumn{3}{l}{Number of Bright Components}\\\vspace{3pt}
{$N_c = 1$   }& 7 & 2 \\
{$N_c = 2$   }& 3 & 2 \\\vspace{5pt}
{$N_c \geq 3$}& 1 & 9 \\
\multicolumn{3}{l}{Mean comp. separation} \\
{$N_c = 2$   }& 8kpc  & 6kpc \\\vspace{5pt}
{$N_c \geq 3$}& 15kpc &18kpc \\
\multicolumn{3}{l}{Physical Extent} \\
\multicolumn{1}{r}{D $\leq$ 10 kpc}&    4 & 3\\
\multicolumn{1}{l}{10 kpc $<$ D $\leq$ 20 kpc}& 4 & 1\\
\multicolumn{1}{l}{20 kpc $<$ D $\leq$ 30 kpc}& 0 & 4\\
\multicolumn{1}{l}{30 kpc $<$ D}&       3 & 5\\
\end{tabular}
\end{center}
\end{table}

Our results (see also Table 5) can be summarised as follows:
\begin{enumerate}
\item[$\bullet$] Seven 6C sources have a single bright component,
three have two bright components\footnote{If the unresolved feature
adjacent to the host galaxy of 6C1256+36 is a foreground star, this
gives eight objects with $N_C = 1$, and only two with $N_C = 2$.}, but
only one source displays more than one additional bright feature.   
\item[$\bullet$] Conversely, only two of the thirteen 3CR sources have
a single bright component, and two have a second bright component. The
remaining nine 3CR sources display three or more bright features.  
\item[$\bullet$] The mean separation of bright components for galaxies
with $N_c = 2$ is less than 10kpc on average in both samples. 
\item[$\bullet$]  If more than one additional
bright feature is observed, the average distance of the bright
components from the galaxy is nearly twice as high, at $\approx 15$kpc
for the 6C sample, and $\approx 18$kpc for the 3CR sample. 
\end{enumerate}
 
Another useful parameter developed in Best (1996) is the {\it
component alignment strength}, $a_c$.  This parameter not only provides a
measure of the degree of alignment, but also the flux in the aligned
features, and particularly the importance of
bright knots in the aligned emission. The component alignment strength is
given by:
\begin{equation}
a_c = N_c a_s = N_c \epsilon \left( 1 - \frac{\Delta \theta}{45} \right),
\end{equation}
where $N_c$ is the number of bright emission components.  
 
\begin{figure}
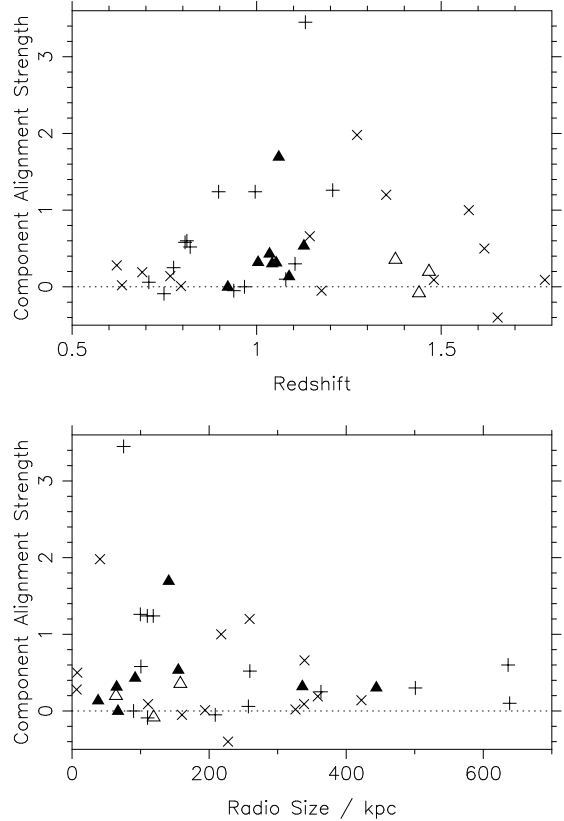

\vspace{4.2 in}
\begin{center}
\includegraphics{ac_d.ps}
\includegraphics{ac_z.ps}
\end{center}
\caption{Component Alignment Strength vs. redshift (top) and radio
size (bottom) for
the 6C and 3CR radio galaxies at $z \sim 1$. Symbols are as in
Fig.~\ref{Fig: 7_6}.
\label{Fig: 7_7}}
\end{figure}

The component alignment strengths for the 6C and 3CR galaxies are
plotted as a function of redshift and radio power in Fig.~\ref{Fig: 7_7}. Once
again, similar trends are observed for both the 6C and 3CR galaxies,
with a peak in the component alignment strength at $z \sim
1.1$.   The slight decrease beyond this redshift is due to the
decreasing sensitivity of the higher redshift observations.
Any anti-correlation with radio size is weaker, and is statistically
insignificant.  We again see a strong ($> 99.5$\% significant)
correlation between component alignment strength and the excess UV flux in our
residual images (again excluding 6C1129+37).
The component alignment strengths of the 6C sources are
generally much lower than those of the 3CR sources, and where a value
of $N_C > 1$ is observed in the 6C sample, this is not so clearly
biased towards smaller radio sources as is seen for the 3CR sources.
Multiple bright knots of emission are rarer in the 6C data (Table 5): at $z \sim 1$ the
average number of bright components is 1.7 for the 6C sample, and 2.7
for the 3CR sample.

Although 6C and 3CR sources at the same redshift have similar values
for the measured strength of the alignment effect, the measured
component alignment strengths were generally less for the 6C
sample. The extended structures of the 3CR sources are clearly more
``knotty'' on average than 6C sources of a similar radio size.  
It is also interesting that where large number of discrete components are
seen, their spatial distributions are similar in both samples,
regardless of radio power.  

\subsection{The alignment effect at $z \sim 1$}

\begin{figure}
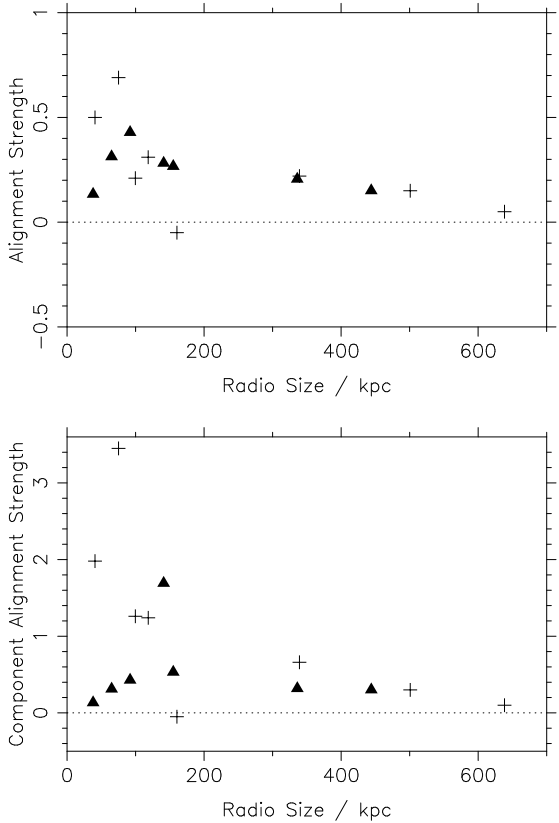

\vspace{4.2 in}
\begin{center}
\includegraphics{az_z.ps}
\includegraphics{az_d.ps}
\end{center}
\caption{(top) Alignment Strength vs. radio size and (bottom) Component
Alignment Strength vs. radio size (for sources with $1.0 < z <1.3$
only). 6C sources are represented by the triangles, and 3CR sources by
the crosses.
\label{Fig: 7_8}}
\end{figure}

The differences between the 6C and 3CR samples become clearer when we
consider a restricted redshift range.  Fig.~\ref{Fig: 7_8} displays
the variation in alignment strength and component alignment strength
with radio size, for the sources in the redshift range $1.0 < z <
1.3$.  At these redshifts, the alignment effect has generally reached
its maximum observed strength, but the data do
not yet suffer from high redshift sensitivity problems.
The alignment strengths for the 3CR data are anti-correlated with radio size at a
significance level of $> 95\%$; adding the 6C data increases this to
$> 97.5\%$.  Whilst both the 6C and 3CR sources show a very
similar result for the variation in $a_s$ with radio size, the results
for $a_c$ clearly show that the smaller 6C sources lack the knotty
structures observed in the 3CR sample.  The 3CR sources show a 99\%
significant anti-correlation between $a_c$ and radio size; the strength
of this correlation is reduced to $\sim 95\%$ with the inclusion of the
6C data points.

If the emission which produces the bright knots of emission is due to
line or nebular continuum emission, the more powerful 3CR sources
should indeed display these features more frequently.  However, an
analysis of the data for two of the sources in the 6C subsample shows
that whilst the diffuse 
extended emission surrounding 6C0943+39 is dominated by line emission,
only a small fraction of the bright knotty emission surrounding
6C1129+37 can be attributed to such processes.  This result implies
that other mechanisms are responsible for the bright knotty emission
predominantly observed around the smaller radio sources. 
The evolution with radio source size of the 3CR aligned
emission has previously been interpreted as being consistent with the effects of
the aging of a young stellar population, whose formation was triggered
by the passage of the expanding radio source (Best, Longair \&
R\"{o}ttgering 1996).
Should this indeed be the case, the morphological differences between the aligned
emission in the 6C \& 3CR subsamples suggests that the drop in radio
power has implications for the efficiency and/or outcome of these processes. 

\begin{table}
\caption{Significance levels of correlations between extended emission
line region size and either alignment strength ($a_s$) or component
alignment strength ($a_c$).  The results are presented for
calculations using both the observed (top) and corrected (bottom) 6C
extended emission line
  region sizes. }
\begin{center} 
\begin{tabular} {ccc} 
 Correlation using observed 6C EELR sizes & $a_s$ &$a_c$ \\\hline
\vspace{0.5pt} 
Full data& $>90$\% & $>95$\%\\\vspace{2.5pt}
Data with $1.0 < z < 1.3$ & $>95$\% & $>95$\%\\\\
Correlation using corrected 6C EELR sizes & $a_s$& $a_c$ \\\hline
Full data & $>95$\% & $>99.5$\%\\
Data with $1.0 < z < 1.3$ & $>99$\% & $>99$\%\\
\end{tabular}
\end{center}
\end{table}

It is particularly useful to contrast our imaging  with the results of
our spectroscopic observations  of the 6C and 3CR $z \sim 1$ subsamples
(Best, R\"{o}ttgering \& Longair 2000a,b;  Inskip et al 2002b,c).  To
summarize, the spectra of these sources show that the
ionization state and kinematic properties of the gas in the extended
emission line regions (EELRs) vary independently with both radio power
and size.  The EELRs of smaller radio sources are more luminous and
extensive, and are best explained by a contribution from shock
ionization associated with the expanding radio source as well as from
AGN photoionization.  Partial rank analysis of 6C and 3C spectra
(Inskip et al 2002c) showed that the gas kinematics are also more extreme
for the sources at higher redshifts, independently of the effects of radio
source power and size/age.  These results suggest that there is some evolution
of the local intergalactic medium (IGM), and that either the distribution and density of gas
clouds varies with redshift, or that interactions/shocks between the IGM and the radio
source are less important at later cosmic epochs.  Whilst the
alignment effect is often observed around lower redshift 
sources, it is generally less 
extreme than that surrounding high-$z$ sources of similar radio
power (Allen et
al 2002, Dey \& van Breugel 1993), indicating 
clear evolution in the mechanisms which produce the alignment effect
as well as the expected dependence on radio power.

\begin{figure}
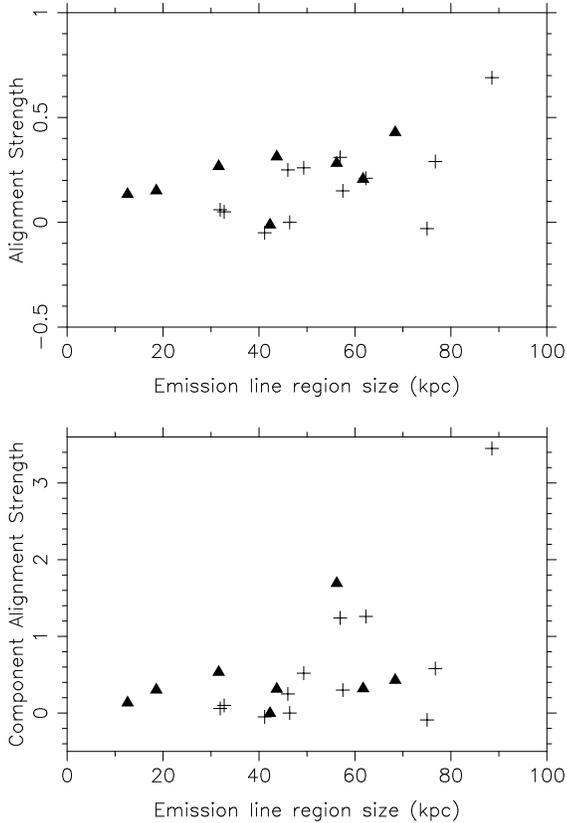

\vspace{4.2 in}
\begin{center}
\includegraphics{eelr_align1a.ps}
\includegraphics{eelr_align2a.ps}
\end{center}
\caption{(top) Alignment Strength vs. extended emission line region size and (bottom) Component
Alignment Strength vs. extended emission line region size (for sources
in the spectroscopic samples of Inskip et al (2002b) and Best et al
(2000a,b). 6C sources are represented by the triangles, and 3CR sources by
the crosses.
\label{Fig: 7_9}}
\end{figure}

The properties of the EELRs seem to be closely linked to the alignment
effect.  The small radio sources in both samples which do not show a
strong UV alignment effect are usually those which have passive,
quiescent emission line regions.  However, where shocks are seen to
boost the emission line luminosity, kinematics and overall extent of
the emission region, we also find that these sources often display a strong
alignment effect. 
Although the extended line emission does not account for a large
proportion of the total flux observed in the rest frame UV (typically
a few percent up to a maximum of 20\% in some sources), the sources
which display the most extreme alignment effect also have the largest
emission line regions: a feature which is itself closely linked to the
importance of shocks (Fig.~\ref{Fig: 7_9}).   Both $a_s$ \& $a_c$ are
correlated 
with emission line region size (details in Table 6).  These
correlations are strengthened if we either restrict the
data to the matched set of objects lying between $1.0 < z < 1.3$, or
if we  scale the 6C emission line region 
sizes upwards by 25\% to reflect the lower sensitivity of the
spectroscopic observations (Inskip et al 2002b).  There is a clear
change between the sources which show a strong alignment effect when
small and young, to the more passive evolved sources that they will
become at later times. This trend for alignment strength and
component alignment strength in particular to be correlated with emission line
region size is actually stronger than that with radio size.
Despite the fact that the trends with radio 
size for the alignment effect are weaker for the 6C sample, this is a
good indication that the underlying trends observed for the 3CR
galaxies still hold true at lower radio powers, and that the presence
of shocks in the extended emission regions may be a driving factor in
producing the most extreme alignment effects.  

\section{Discussion}

The analysis of the morphologies and alignment strengths of
$z \sim 1$ 6C and 3CR radio galaxies has led to a number of
interesting results.
\begin{enumerate}
\item[$\bullet$] The 6C host galaxies at $z \sim 1$ are best described
as de Vaucouleurs ellipticals.  Fitting with S\'{e}rsic profiles gives
a mean value of $n \sim 4$ for the sample; the range of values
determined for $n$ ($2 \lta n \lta 6$) is comparable to that
found by Zirm, Dickinson \& Dey 
(2003) in their similar analysis of 3CR sources at $z \sim 1$. It is
noteworthy that the two sources which have $n > 4$ are amongst 
the three most luminous in the sample; larger values of $n$ are
frequently observed to be correlated with the mass/luminosity of the
host galaxy (e.g. Graham \& Guzm\'{a}n 2003). 
\item[$\bullet$]Typically, the 6C sources have effective radii of $\sim
9-10$kpc and an average unresolved nuclear point source contribution of
16\% in the $K$-band.  The values of these parameters are comparable
to those of low redshift radio galaxies matched in radio power, and
also to more powerful radio sources at the same redshift.
\item[$\bullet$] The measured alignment strengths, and the variations
in this parameter with both redshift and radio size are almost
identical for both the 6C and 3CR $z \sim 1$ subsamples.  Both $a_s$ and $a_c$ peak at a redshift
of $z \approx 1.1$, and are anti-correlated with radio source
size. Similar ellipticities are measured for the smoothed images, and
the extended structures are equally well aligned with the radio axis.
\item[$\bullet$] On average, the component alignment strength is lower
for the 6C sources.  These display fewer bright components, and their
aligned structures are generally less extensive.
\item[$\bullet$] However, for sources matched in radio power, the higher redshift sources
display a stronger alignment effect.  This reflects the
decreasing importance of interactions at low redshifts, and also the possible
evolution of the distribution of material in the surrounding IGM. 
\item[$\bullet$] The observed trend for the smaller 3CR sources to
display brighter, more extensive and better aligned morphologies (Best
{\it et al} 1996) is clearly not as strong for the galaxies in the 6C
subsample, which generally have less extreme morphologies.
\item[$\bullet$] Regardless of radio size, the sources which display an extreme
alignment effect also have the most extensive emission regions; these
are the sources for which shocks are most important. 
\end{enumerate}
\vspace{-0.2cm}

Our results add further weight to the picture that the various features of
the alignment effect depend strongly on both redshift and radio source
power.  
Radio power remains an important factor at lower redshifts, as is
evident from the study of a sample of
still less powerful 7C radio galaxies at $z \sim 0.7$ (Lacy et al
1999). For the 7C systems, aligned emission is still observed, but is
substantially weaker than that of the matched sample of more powerful
3CR sources at the same redshift. The percentage aligned flux is found
to be lower (7\% for 7C cf. 18\% for 3C, at wavelengths slightly
longer than the 4000\AA\ break), and the excess emission exists on 
smaller spatial scales than for the higher radio luminosity 3CR sources
at the same redshift. Additionally, neither sample displays any trends between
alignment effect and radio source size; the luminous clumpy features most likely
to be responsible for the radio size trends observed in the 3CR $z
\sim 1$ subsample are far less common at lower redshifts and/or radio powers. 
In terms of evolution with redshift, the 6C alignment effect at $z \sim 1$ is
more frequent and noticeably stronger than that observed for galaxies of a similar radio
luminosity at lower redshift.  Dey \& van Breugel (1993) observed significant UV
alignment effects in only 30\% of low redshift systems; similar
results are seen in other low redshift studies (e.g. de Koff et al 1996).

Given that the alignment effect clearly varies in strength with radio
source power as well as redshift, one might expect its behaviour to be predictable.  Of
the various mechanisms proposed to explain the alignment effect,
several should scale with radio power.  [O\textsc{ii}] emission and
the bulk kinetic power of the radio jet are both observed to scale
with radio source luminosity to the power of $\sim$6/7, over a wide
range of source powers and redshifts (Rawlings \& Saunders 1991;
Willott et al 1999).  Scattered AGN emission and total line emission
are therefore also expected to increase with the power of the radio
source.  On the other hand, star formation induced by the passage of
the radio source jets 
may  be favoured in the case of lower power radio sources (perhaps
explaining the strong alignment effects observed in the
case of the high redshift 6C sources), where the bulk kinetic
energy carried by the jets is correspondingly lower (Lacy et al
1999).  However, this does not rule out star formation induced by
lower velocity bulk motions associated with backflows or the sideways expansion of
the radio cocoon in the case of more powerful sources.  Given the
strong alignment effects observed for both $z \sim 1$ subsamples, the optical/UV
alignment effect clearly depends on a number of other factors in
addition to radio power, which themselves become
unable to induce such a strong alignment effect at lower redshifts. 

The emission at longer wavelengths is also worthy of further discussion.  
The lack of any IR alignment effect in our 6C subsample suggests that
the long wavelength tail of the alignment effect does indeed scale
most strongly with radio power.   However, the long-wavelength
alignment effect cannot account for differences in $K$-band emission
between the 6C and 3CR samples, nor is it sufficient to explain the
fact that the host galaxies of high redshift radio sources are more luminous than expected based on
the predictions of passive evolution models (Inskip et al 2002a).  The
strength of any nuclear point source contribution cannot be
responsible, as the point source percentage in the $K$-band is similar
for both $z \sim 1$ samples, regardless of radio power.   Variations
in $K$-band luminosity cannot easily be explained by host galaxy mass
either; the 6C host galaxies are massive ellipticals very similar in
size to those of other radio galaxy samples matched in either radio
power or redshift. Below $z \sim 2$, the radio source host galaxy population appears
remarkably uniform, with scale sizes in the region of 10kpc: well
below those of brightest cluster galaxies, but still larger than is
typical for normal low redshift ellipticals.   This is in good
agreement with the apparent passive evolution of the lower power radio
sources in the $K$-band magnitude--redshift (discussed in detail in
Inskip et al 2002a).  

The only remaining possibility is a significant proportion of young stars in
the host galaxies of the more powerful sources.  The problem with this
approach is that the presence of a young stellar population of $\lta
10^8$years in age is itself constrained by the shorter wavelength
galaxy colours to a fairly limited proportion of the host galaxy total
stellar mass ($< 5$\%, equivalent to a maximum 0.5mag increase in
$K$-band luminosity; BLR98 \& Inskip et al 2002a).  However, recent
research (e.g Tadhunter et al 2002; Wills et al 2002; Tadhunter et al
2005) has identified the clear presence of older (0.1-2Gyr), often
reddened, young stellar populations, which could provide an increase
in the observed IR emission without overly boosting the rest-frame UV
beyond the observational limits. Such a scenario would also fit in
well with the recent results from the Sloan Digital Sky Survey, which
found that the highest luminosity AGN were hosted by galaxies with 
relatively young stellar populations (Kauffmann et al 2003). 

We have investigated the influence of un-reddened young stellar populations on
the galaxy emission.  Using the \textsc{GISSEL} spectral
synthesis models of Bruzual \& Charlot (2003), we find that in order
to produce a 0.5mag increase in $K$-band magnitude whilst avoiding
large increases in the rest-frame UV emission from the host galaxy,
the young stellar 
population is required to contain $\sim 20$\% of the host galaxy mass and
be over 0.5Gyr in age, i.e. significantly older than the radio
source.  Such an age could argue against any direct link with radio
source power, and it is not obvious that this could explain any
$K$-band magnitude difference between the 6C and 3CR data. However,
if such a starburst was associated with a merger event which led to
the subsequent triggering of the radio source/AGN activity, then the
exact details of that merger could provide a link between any young
stellar population and the power of the resulting radio source.  
It is also possible that the more diffuse aligned emission (which displays
less radio power dependence) could be associated with such older,
merger-related starbursts. West (1994) suggests that mergers occur
anisotropically and that the radio source jets are likely to be
orientated in alignment with the surrounding matter distribution,
thereby leading to an alignment with a young stellar population which
could be older than the radio source itself.  On top of this, other
processes (including later radio source/jet-induced
star formation) would be responsible for producing the more luminous,
clumpy aligned emission, which does show a strong dependence on both
radio source power and age.
In addition to a strong UV excess, such very recent star formation 
($10^7$ to $5 \times 10^7$ years in age) could
also generate an increase in flux at longer wavelengths sufficient to
explain the fact that these galaxies are typically 0.5mag brighter than the 
predictions for purely passively evolving systems at $z \sim 1$
(Inskip et al 2002a).   However, for consistency with our observed
data, for most sources such a young starburst would be limited to a
few percent of the total galaxy mass, and would need to be reddened
by an E(B-V) of typically 0.2-0.6.  These issues will be considered in
greater depth in Paper III of this series.  

We expect that the most likely solution for explaining the evolution
of the $K$-band galaxy properties is a combination of several factors: 
smaller variations in galaxy mass, recent star formation and
contamination by processes associated with the radio source
activity/alignment effect.  At higher redshifts, it is likely that
increased merger activity goes hand in hand with the presence of
powerful radio galaxies and ongoing star formation (e.g. studies of
ultraluminous infrared galaxies: Genzel et al 2001; Tacconi et al
2002; Bushouse et al 2002).  The morphologies of high redshift ($z >
3$) radio sources are clearly disturbed, multi-component systems,
whilst at slightly lower redshifts ($z \sim 2$) although the galaxies
give the appearance of settled elliptical systems, their typical scale
sizes ($\sim 6$kpc) are very much smaller than those of radio sources
at $z \sim 1$ (Pentericci et al 1999; van Breugel et al 1998),
suggesting that they have not yet completed their process of assembly.
It is certainly plausible that such  processes are still affecting the
lower redshift $z \lta 1$ systems on a smaller scale, but more so in
the case of the most luminous radio sources such as the $z \sim 1$ 3CR
galaxies, which also display the most extreme examples of the
alignment effect.

\section*{Acknowledgements}
We would like to thank the referee, Mark Lacy, for several very useful
comments.
KJI acknowledges the support of a
Lloyds' Tercentenary Foundation Research Fellowship and a PPARC
Postdoctoral Research Fellowship.  PNB is
grateful for the generous support offered by a Royal Society Research
Fellowship. The United Kingdom Infrared Telescope is operated by the
Joint Astronomy Centre on behalf of the U.K. Particle Physics and
Astronomy Research Council. Some of the data reported here were
obtained as part of the UKIRT Service Programme.
Parts of this research are based on observations made with the
NASA/ESA Hubble Space Telescope, obtained at the Space Telescope
Science Institute, which is operated by the Association of
Universities for Research in Astronomy, Inc., under NASA contract NAS
5-26555. These observations are associated with proposals \#6684 and
\#8173.

\label{lastpage}

\end{document}